\documentclass[aps,prb,floats]{revtex4}

\usepackage{ifthen}
\usepackage{ifpdf}

\usepackage{latexsym}
\usepackage{amsmath}
\usepackage{amssymb}
\usepackage{bm}
\usepackage{wasysym}

\ifpdf
\usepackage{graphicx}
\usepackage{epstopdf}
\else
\usepackage{graphicx}
\usepackage{epsfig}
\fi

\newcommand{\be}{\begin{equation}}
\newcommand{\ee}{\end{equation}}

\newcommand{\ep}{\epsilon}

\def \a{\vec a}
\def \A{\vec A}
\def \K{\vec K}
\def \G{\vec G}
\def \D{\vec D}
\def \q{\vec q}
\def \R{\vec R}
\def \r{\vec r}
\def \q{\vec q}

\begin{document}

\title{A universal Hamiltonian for the motion and the merging of Dirac cones \\ in a
two-dimensional crystal}

 \author{G. Montambaux,  F. Pi\'echon, J.-N. Fuchs and  M. O. Goerbig}
\affiliation{Laboratoire de Physique des Solides, CNRS UMR 8502,
Universit\'e Paris-Sud, 91405- Orsay, France}
\date{\today}

\begin{abstract}
We propose a simple Hamiltonian to describe the motion and the merging of Dirac points in the electronic
spectrum of two-dimensional electrons. This merging is a topological transition which separates
a semi-metallic phase with two Dirac cones
 from an insulating phase with a gap. We calculate the density of states and  the specific heat. The
 spectrum in a magnetic field $B$ is related to the resolution of a
 Schr\"odinger equation in a double well potential. They obey the general scaling law
 $\ep_n \propto B^{2/3} f_n(\Delta /B^{2/3}$. They
 evolve continuously from a $\sqrt{ n B}$
 to a linear $(n+1/2)B$ dependence, with a  $[(n+1/2)B]^{2/3}$ dependence at the transition. The spectrum in the vicinity of the topological transition
 is very well described by a semiclassical quantization rule. This model describes
 {\it continuously} the coupling between valleys associated with the two Dirac points, when approaching the transition.
 It is applied to the tight-binding model of graphene and its generalization when one hopping parameter
 is varied. It remarkably reproduces the low field part of the Rammal-Hofstadter spectrum for
  the honeycomb lattice.
\end{abstract}

\maketitle

\section{Introduction}

The main interest of graphene from the fondamental
point of view is that the low energy electronic spectrum (around the band center $\ep=0$)
is linear, exhibiting the so-called Dirac spectrum around two special
points $\K$ and $\K'$ at the corner of the Brillouin zone (BZ).\cite{Wallace}
As a consequence, the density of states  varies linearly with energy. In a magnetic field $B$,
the energy levels around $\ep=0$ vary  as $\ep_n(B) \propto \sqrt{n B}$, with a two-fold degeneracy
corresponding to the two valleys near $\K$ and $\K'$. The   considerable development of research on graphene
is partly due to this unusual spectrum.\cite{review}

The electronic spectrum of graphene is very well described by a tight binding model on a honeycomb lattice, with three
equal couplings $t$ between nearest neighbors.\cite{Wallace} It has been soon realized that, by varying
these hopping parameters,  new interesting physics could emerge, in particular the existence of a topological
transition separating  a metallic phase with two Dirac points and an insulating phase with a
 gap.\cite{Hasegawa1,Dietl,CastroNeto2,Guinea,Montambaux09,Segev,Volovik}. When only one of the three hopping parameters is modified ($t'$, see
 figure \ref{fig.reseau}), the transition
occurs when $t'=2 t$. This model that we will call the $t-t'$ model is generic and contains the essential physics of the more general case when the three hopping integrals are different.

Although such variation of hopping parameters may not be feasible in graphene,  a transition
could well be observed in other systems like the organic conductor $\alpha-(BEDT-TTF)_2I_3$
 \cite{Katayama2006,Kobayashi2007,Goerbig2008} or an artificial lattice of cold atoms,\cite{Zhu,Zhao,Hou2009,Lee09} where the motion of Dirac points
 may be induced by changing the intensity of the laser fields.

The spectrum in a magnetic field of the honeycomb lattice has been first considered at low field by McClure \cite{McClure} who
found a $\sqrt{n B}$ dependence of the energy levels near the band center, and   R. Rammal  calculated
the so-called "Hofstadter" spectrum which describes the fractal broadening of the Landau levels, due the
the competition between magnetic field and lattice effects.\cite{Hofstadter,Rammal} More recently the Hofstadter-Rammal spectrum
has been studied when hopping parameters are modified,\cite{Hasegawa2} and it was found that at the topological transition
the low field spectrum exhibits a new dependence with the magnetic field of the form
 $[(n+1/2) B]^{2/3}$.\cite{Dietl} This is due to the  peculiar character of the dispersion relation
 at the transition: it is linear in one direction and quadratic in the other one.
  Quite recently it has been proposed that such a hybrid spectrum and the subsequent structure of the Landau levels may  exist
  in $VO_2/TiO_2$ nanostructures.\cite{Banerjee}

In a recent paper, we have studied under which general conditions a pair of Dirac points in the electronic spectrum
of a two-dimensional crystal merges into a single point at the topological transition.\cite{Montambaux09} We have derived
a low energy Hamiltonian that describes the physical properties near the transition. The present paper considers in details these
physical properties. The main interest of this work is to present
the simplest model which continuously describes the merging of two Dirac points and the subsequent gap opening,  and to calculate several physical properties in the vicinity of the transition.
We wish to stress that  the interest of this work is not only
to tune continuously the coupling between two Dirac valleys, but also to study the combination between two
distinct and quite interesting dispersion relations : the linear dispersion relation and a dispersion relation with
a saddle point. Indeed, the two Dirac points are always separated by a saddle point, and the merging of Dirac points is obviously
accompanied by the merging with the saddle point. At this merging, a linear density of states characteristic of the
Dirac point approaches a logarithmic density of states characteristic of a saddle point (in $2D$).

The paper is organized as follows. In the next section, we consider a general tight binding problem in $2D$, with
two atoms per unit cell and study under which general conditions Dirac points may merge. In section III, we study several
properties of this Hamiltonian. In particular, we relate the Landau level spectrum to a one-dimensional double well problem, where the two
wells correspond to the two valleys around the Dirac points. When approaching the transition, the potential well vanishes and
the potential becomes quartic at the transition. In section IV, we show how to relate the parameters of the universal Hamiltonian to the parameters
$t$ and $t'$ of the tight binding model on the honeycomb lattice, the so-called $t-t'$ model.
In this way, we reproduce with a very good accuracy the low field part of the
butterfly spectrum spectrum and the lifting of the valley degeneracy of the Landau levels

\section{Construction of the universal Hamiltonian}

We consider a  two-band  Hamiltonian for a $2D$
crystal with two atoms $A$ and $B$ per unit cell.
This two-band Hamiltonian is naturally given in terms of the $2\times 2$ matrix
$$ {\cal H}({\vec k})= \left(%
\begin{array}{cc}
h_{AA}({\vec k})  & h_{AB}({\vec k}) \\
  h_{BA}({\vec k}) &  h_{BB}({\vec k}) \\
\end{array}%
\right) \ ,  $$ with the 2D wave vector ${\vec k}$. Time-reversal
symmetry (${\cal H}({\vec k})={\cal H}^*(-{\vec k})$) imposes
${h_{AB}}({\vec k})={h_{BA}}^*({\vec k})\equiv f({\vec k})$ and,
together with hermiticity, real symmetric diagonal terms
$h_{AA}({\vec k})=h_{AA}(-{\vec k})$ ($h_{BB}({\vec
k})=h_{BB}(-{\vec k})$). Furthermore, we consider a 2D lattice with
inversion symmetry such that $h_{AA}({\vec k})=h_{BB}({\vec k})$.
The resulting energy dispersion reads $\epsilon_{\pm}({\vec
k})=h_{AA}({\vec k})\pm |f({\vec k})|$, and we will even set
$h_{AA}({\vec k})=0$ because this term simply shifts the energy as a
function of the wave vector but does not affect the topological
properties of the semi-metal-insulator phase transition discussed
below.

We, therefore, discuss from now on the Hamiltonian in its reduced form

\be {\cal H}({\vec k})= \left(%
\begin{array}{cc}
0  & f({\vec k}) \\
  f^*({\vec k}) &  0 \\
\end{array}%
\right) \ , \label{H} \ee where the off-diagonal terms have the
periodicity of the Bravais lattice and may be written quite
generally in the form:

\be f({\vec k})= \sum_{m,n} t_{mn} e^{- i {\vec k} \cdot \R_{mn}} \
, \label{fofk} \ee
where the $t_{mn}$'s are real, a consequence of time-reversal
symmetry ${\cal H}({\vec k})={\cal H}^*(-{\vec k})$, and   $\R_{mn}=
m \a_1 + n \a_2$ are vectors of the underlying Bravais lattice.

The energy spectrum is given by  $\ep({\vec k})=\pm |f({\vec k})|$,
and the Dirac points, that we name $\D$ and $-\D$ are solutions of
$f(\D)=0$. Since $f({\vec k})=f^*(-{\vec k})$, the Dirac points,
when they exist, necessarily come in by pairs.\cite{remark2} The
position $\D$ of the Dirac points can be anywhere in the BZ and move
upon variation of the band parameters $t_{mn}$. Around the Dirac
points $\pm \D$, the function $f({\vec k})$ varies linearly. Writing
${\vec k} = \pm D+\q$, we find

\be f(\pm \D+\q)=\q \cdot (\pm {\vec v}_1 - i {\vec v}_2)
 \label{fofq}
\ee
where the velocities ${\vec v}_1$ and ${\vec v}_2$ are given by
\begin{eqnarray}
{\vec v}_1 &=& \sum_{mn} t_{mn} \R_{mn} \sin \D \cdot \R_{mn} \nonumber \\
{\vec v}_2 &=& \sum_{mn} t_{mn} \R_{mn} \cos \D \cdot \R_{mn}
\label{v1v2}
\end{eqnarray}

Upon variation of the band parameters, the two Dirac points may
approach each other and merge into a single point $\D_0$. This
happens when $\D=-\D$   modulo a reciprocal lattice vector $\G=p
\a^*_1 + q \a_2^*$, where  $\a^*_1$ and  $\a_2^*$ span the
reciprocal lattice. Therefore, the location of this merging point is
simply $\D_0= \G/2$. There are then four possible inequivalent
points the coordinates of which are $\D_0= ( p\a^*_1 + q \a_2^*)/2$,
with $(p,q)$ = $(0,0)$, $(1,0)$, $(0,1)$, and $(1,1)$.
 The condition
$ f(\D_0)= \sum_{mn} (-1)^{\beta_{mn}} t_{mn} =0$, where
$\beta_{mn}= p m + q n$, defines a manifold in the space of band
parameters. As we  discuss below, this manifold separates a
semi-metallic phase with two Dirac cones and a band insulator.

In the vicinity of the $\D_0$ point, $f$ is {\it purely imaginary}
(${\vec v}^0_1=0$), since $\sin (\G \cdot \R_{mn}/2)=0$.
Consequently, to lowest order, the linearized Hamiltonian  reduces
to ${\cal H}= \q
 \cdot {\vec v}_2^0  \sigma^y$, where  ${\vec v}^0_2=\sum_{mn} (-1)^{\beta_{mn}} t_{mn} \R_{mn} $.
We choose the local  reference system such that ${\vec v}^0_2 \equiv
c_y \, \hat y$ defines the $y$-direction.\cite{directions} In order
to account for the dispersion in the local $x$-direction,  we have
to expand $f(\D_0+\q)$ to second order in $\q$:
 \be f(\D_0+\q)= - i \q \cdot {\vec v}^0_2 -{1 \over 2} \sum_{mn} (-1)^{\beta_{mn}}
  t_{mn} ( \q \cdot
 \R_{mn})^2 \ .  \label{fofq0} \ee
Keeping the quadratic term in $q_x$, the new Hamiltonian
 may be written as
\be {\cal H}_0(\q)   = \left(
  \begin{array}{cc}
    0 & {q_x^2 \over 2 m^*} -  i c_y q_y  \\
 {q_x^2 \over 2 m^*} +  i c_y q_y & 0 \\
  \end{array}
\right) \ .  \label{H0topo}\ee
where the effective mass $m^*$ is defined by
 \be {1 \over m^*}=  \sum_{mn} (-1)^{{\beta_{mn}}+1} t_{mn}
 R^2_{mn,x} \ ,  \ee
and where  $R_{mn,x}$ is the component of $\R_{mn}$ along   the {\it
local} $x$-axis (perpendicular to ${\vec v}_2^0$).  The terms of
order $q_y^2$ and $q_x q_y$ are neglected at low energy.
  The diagonalization of ${\cal H}_0(\q)$ is
straightforward and the energy spectrum

\be \ep = \pm \left[ c_y^2 q_y^2+\left({q_x^2 \over  2 m^*}\right)^2\right]^{1/2}
\ee
 has a remarkable structure: it
is linear in one direction and quadratic in the other.   From the
linear-quadratic spectrum which defines a velocity $c_y$ and a mass
$m^*$, one may identify a characteristic energy :

\be m^* c_y^2= {[\sum_{mn} (-1)^{\beta_{mn}} t_{mn} \R_{mn} ]^2 \over
\sum_{mn} (-1)^{{\beta_{mn}}+1} t_{mn} R^2_{mn,x} } \ . \ee

The merging of the Dirac points in $D_0$ marks the transition
between a semi-metallic phase and an insulating phase. In this
paper, we concentrate on the properties of the spectrum in the
vicinity of the merging. The transition is driven by the parameter

\be \Delta= f(\D_0)= \sum_{mn} (-1)^{\beta_{mn}} t_{mn} \label{gap} \ee
which changes  its sign at the transition. This parameter $\Delta$
therefore drives the transition. In the vicinity of the transition,
the Hamiltonian has the form

\be {\cal H}(\q)= 
\left(
  \begin{array}{cc}
    0 & \Delta+ {q_x^2 \over 2 m^*} -  i c_y q_y  \\
 \Delta + {q_x^2 \over 2 m^*} +  i c_y q_y & 0 \\
  \end{array}
\right) \label{newH} \ee
with the spectrum
\be \ep= \pm \sqrt{(\Delta + {q_x^2 \over 2 m^*})^2 + q_y^2 c_y^2}
\label{Ueps1} \ee

 The Hamiltonian ({\ref{newH}) has a remarkable structure
and describes properly the vicinity of the topological transition,
as shown on Fig. \ref{fig.bicones}. When $ m^* \Delta$ is negative
(we  choose $m^* >0$ without loss of generality), the spectrum
exhibits the two Dirac cones and a saddle point in $\D_0$ (at half
distance between the two Dirac  points). Increasing $\Delta$ from
negative to positive values, the saddle point evolves into the
hybrid point  at the transition   ($\Delta=0$) before
 a gap $2 \Delta >0$ opens.

In this paper, we study the spectral properties around the merging,
in particular in the presence of a magnetic field. Moreover, we
stress that this Hamiltonian has the general structure to describe
the physics of Dirac points, even far from the transition, since it
captures quite simply the coupling between the two valleys
associated with  the two Dirac points. In particular, we can relate the
coupling between valleys to a double well potential problem. For
this reason we name it a universal Hamiltonian.

\section{Properties of the universal Hamiltonian}

\begin{figure}[!h]
\centerline{ \epsfxsize 7cm  \epsffile{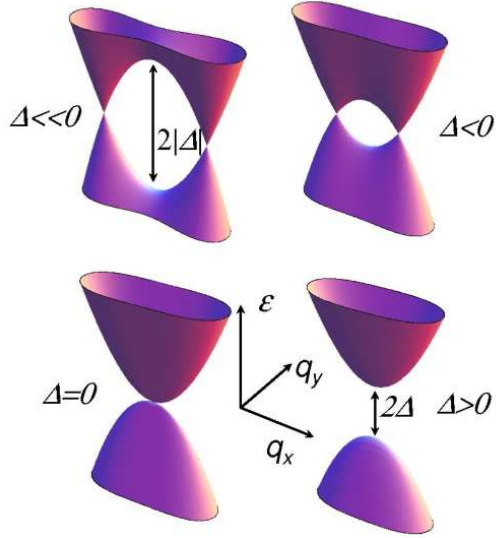}}
\caption{\it Evolution of the spectrum when the quantity $\Delta$ is
varied and changes in sign  at the topological transition (arbitrary
units).  The low-energy spectrum  stays linear in the $q_y$
direction.} \label{fig.bicones}
\end{figure}

Without loss of generality, we assume $m^* >0$. When $\Delta$ varies
from negative to positive values, a topological transition from a
semi-metallic phase with two Dirac cones and a band insulator with a
gapped spectrum occurs. At the transition, the spectrum is hybrid, a
reminiscence of the saddle point in the semi-metallic phase, see
figure (\ref{fig.bicones}).

When $\Delta <0$, the spectrum exhibits two Dirac points the
position of which along the $x$ axis is given by $\pm q_D$ with

\be q_D = \sqrt{- 2 m^* \Delta} \ee
and the linear spectrum around these Dirac points is characterized
by the velocity $c_x$ along the $x$ direction :
\be c_x={q_D \over m^*}=   \sqrt{{- 2  \Delta \over m^*}} \ . \label{cxqDm} \ee
The two Dirac points are separated by a saddle point  at position
$q_S=0$ whose energy is $\pm |\Delta|$. The mass $m^*$ describes the
curvature of the spectrum at the vicinity of this point along the
$x$ direction. When $\Delta <0$ varies, the Dirac points move along
the $q_x$ axis.

The energy dispersion relation (\ref{Ueps1}) is characterized by
three parameters, the velocity $c_y$ along the $q_y$ direction, the
mass $m^*$ along the $q_x$ direction and the gap $\Delta$.
Alternatively, it can be characterized by the distance $2 q_D$
between the Dirac cones or the velocity $c_x$, or by any combination
of two among the four parameters $m^*, \Delta, c_x$, or $q_D$. In
table (\ref{table.parameters}), we write explicitly all the
combinations between these parameters.
\begin{table}
\center{\begin{tabular}{|c|c|c|}
  \hline
  $q_D$, $m^*$ & $c_x=q_D/m^*$ & $\Delta= - q_D^2/2 m^*$ \\
   $q_D$, $c_x$ & $m^*=q_D/c_x$ & $\Delta= -c_x q_D/2$ \\
  $m^*$, $c_x$ & $q_D=m^* c_x$ & $\Delta = - m^* c^2 /2$ \\
 $m^*$, $\Delta$ & $c_x=\sqrt{-2 \Delta /m^*}$ & $q_D= \sqrt{- 2 m^* \Delta}$ \\
  $c_x$, $\Delta$  & $q_D= - 2 \Delta /c_x$ & $m^*= -2 \Delta/c_x^2$ \\
$q_D$, $\Delta$ & $m^*= -q_D^2/2 \Delta$ & $c_x= - 2 \Delta/q_D$ \\
  \hline
\end{tabular}}
\caption{\em In addition to the velocity $c_y$, the universal Hamiltonian is described by two
independent parameters (left column) from which two other parameters
may be deduced}
\label{table.parameters}
\end{table}
The universal Hamiltonian describes properly the
vicinity of the Dirac points. The spectrum can be linearized along
the $q_x$ direction, to recover a anisotropic Dirac equation in the
vicinity of each point with a velocity $c_x$ given by (\ref{cxqDm})

$${\cal H}= \left(
              \begin{array}{cc}
                0 & \pm c_x \delta q_x - i c_y q_y \\
              \pm c_x \delta q_x + i c_y q_y & 0\\
              \end{array}
            \right)
$$
where $\delta q_x= q_x - q_D$.

In section IV, we discuss which combination of parameters should be taken in order to properly describe the low energy physics of
the $t-t'$ model of the honeycomb lattice.

\bigskip

\subsection{Density of states}

We have calculated the density of states for the energy dispersion (\ref{Ueps1}). We find

\begin{eqnarray}
\ep < |\Delta| & \quad \rightarrow \quad & \rho(\ep) = {\sqrt{2 m^*}
 \over \pi^2  c_y} {\ep \over \sqrt{\ep -\Delta}}\
K\left(\sqrt{{2 \ep \over
\ep - \Delta}} \right) \nonumber \\
\ep > |\Delta| & \quad \rightarrow \quad & \rho(\ep) = {\sqrt{2 m^*}
 \over \pi^2  c_y} \sqrt{{\ep \over 2}}  \
K\left(\sqrt{{ \ep -\Delta \over
2 \ep }} \right)  \label{DOS}
\end{eqnarray}
where $K(x)$ is the complete elliptic integral of the first kind.\cite{gradstein} In
the low energy limit, one recovers the familiar linear energy
dependence $\rho(\ep)= {1 \over \pi c_x c_y} \ep$. The density of
states exhibits a logarithmic divergence at $\ep =|\Delta|$, due to
the saddle point. It is plotted in figure (\ref{fig.dos}) for a
fixed mass $m^*$, and upon variation of the parameter $\Delta$. When
approaching the transition, the weight of the logarithmic singularity
vanishes and, at the transition, one recovers the density of states found in ref. \onlinecite{Dietl}, given by

\be \rho(\ep)= C {\sqrt{m^*} \over c_y} \ep^{1/2}    \ee
where $C={1\over \pi^2} K(1/\sqrt{2})=\Gamma(1/4)^2 /(4 \pi^{5/2}) \simeq 0.188$.

Above the transition, there is a finite gap $\Delta >0$ and the density of states has a jump at $\ep=\Delta$ :

\begin{eqnarray}
\ep < \Delta & \quad \rightarrow \quad & \rho(\ep) = 0 \nonumber \\
\ep > \Delta & \quad \rightarrow \quad & \rho(\ep) = {\sqrt{2 m^*}
 \over \pi^2  c_y} \sqrt{{\ep \over 2}}  \
K\left(\sqrt{{ \ep -\Delta \over
2 \ep }} \right) \ .   \label{DOS2}
\end{eqnarray}

\begin{figure}[!h]
\centerline{ \epsfxsize 8cm  \epsffile{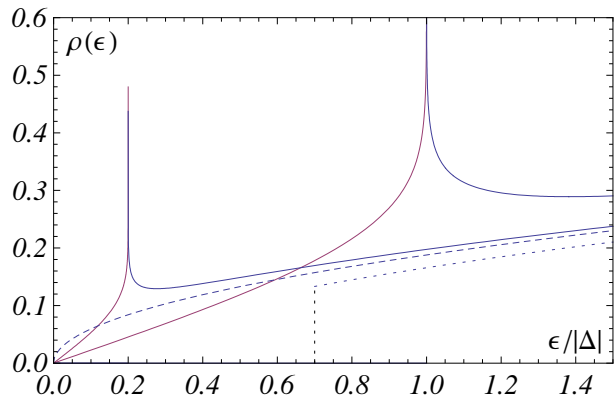}} \caption{\it
Evolution of the density of states on the metallic side of the
transition ($\Delta <0$), plotted for $\Delta=-1$ and $\Delta=-0.2$. For finite $\Delta$, there is a linear
dependence at low energy $\ll |\Delta|$, followed by a logarithmic
singularity at $|\Delta|$. At the transition the density of states
varies as $\sqrt{\ep}$ (dashed curve). Above the transition, there is a finite gap ($\Delta >0$)
and the density of states has a discontinuity (dotted curve for $\Delta=0.7$).} \label{fig.dos}
\end{figure}

\subsection{Specific heat}

Since the spectrum has the electron-hole symmetry $\rho(\ep)=\rho(-\ep)$, the chemical potential is temperature independent
and fixed at $\mu=0$ for the undoped system. Therefore the specific heat has the general form

\be C(T)= {k_B \beta^2 \over 2} \int_0^\infty {\ep^2 \rho(\ep) d \ep \over \cosh^2 {\beta \ep \over 2} } \ee
where $\beta=1/(k_B T)$. Using the expression (\ref{DOS}) of the density of states, we obtain, on the metallic side ($\delta <0$)
\be C(T)= 4 k_B {\sqrt{2 m^*} \over \pi^2 c_y} (k_B T)^{3/2} f\left({T \over |\Delta|}\right) \label{CVf} \ee
where the function $f(T/|\Delta|)$ is given by

\be f(t)= \int_0^{1/2 t} { 2 \sqrt{t}\, x^3 \over \sqrt{2 x t + 1} \cosh^2 x }K(\sqrt{{4 x \over 2 x t +1}}) dx +
\int_{1/2 t}^\infty {   x^{5/2} \over  \cosh^2 x }K(\sqrt{{2 x t+1 \over 4 x }}) dx \ee
It
is plotted on figure (\ref{fig.CV}) and it has the following limits
\bigskip

\begin{tabular}{ll}
                                                                $t \rightarrow 0$ & $f(t) \rightarrow { 9 \pi \over 8 } \zeta(3)\sqrt{t} \simeq 4.248 \sqrt{t}$ \\
                                                                \\
                                                                $t \rightarrow \infty$ & $f(t) \rightarrow {15 \over 256 }(\sqrt{8}-1) \Gamma(1/4)^2 \zeta({5/2}) \simeq 1.889$ \\
                                                                \end{tabular}
\bigskip

\noindent
so that the specific heat interpolates from a $T^2$ behavior far from the transition to a $T^{3/2}$ behavior at the transition. Similarly, above the transition,
in the insulating phase ($\delta >0$), we find, using (\ref{DOS2}):

\be C(T)= 4 k_B {\sqrt{2 m^*} \over \pi^2 c_y} (k_B T)^{3/2} g\left({T \over \Delta}\right) \label{CVg} \ee
where the function $g(T/\Delta)$ is given by

\be g(t)=
\int_{1/2 t}^\infty {   x^{5/2} \over  \cosh^2 x }K(\sqrt{{2 x t-1 \over 4 x }}) dx \ee

\begin{figure}[!h]
\centerline{ \epsfxsize 7cm  \epsffile{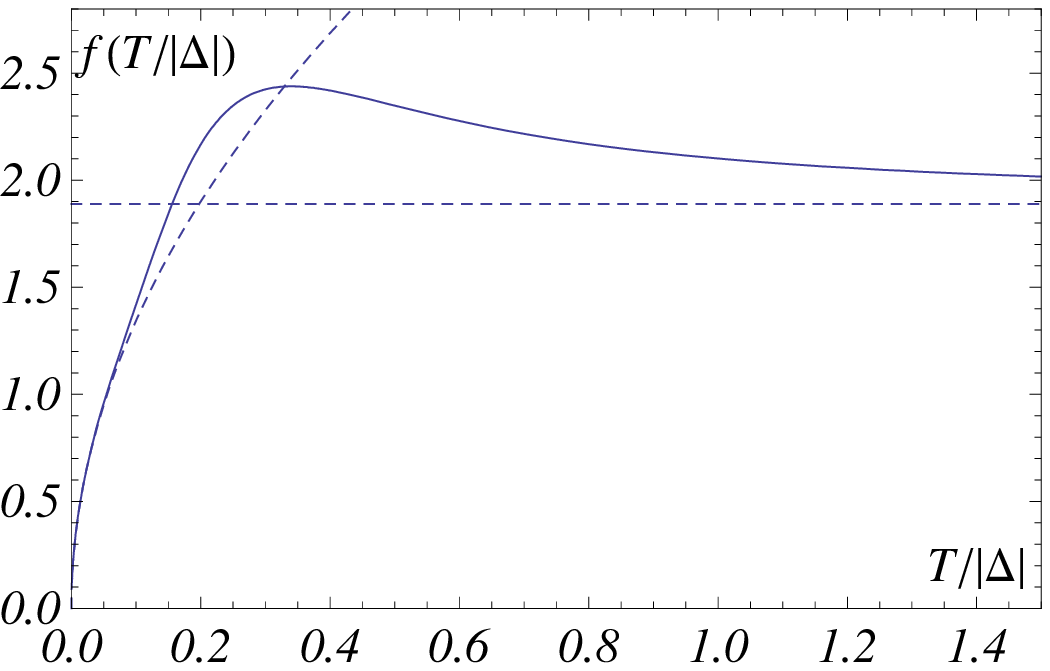}}
\centerline{ \epsfxsize 7cm  \epsffile{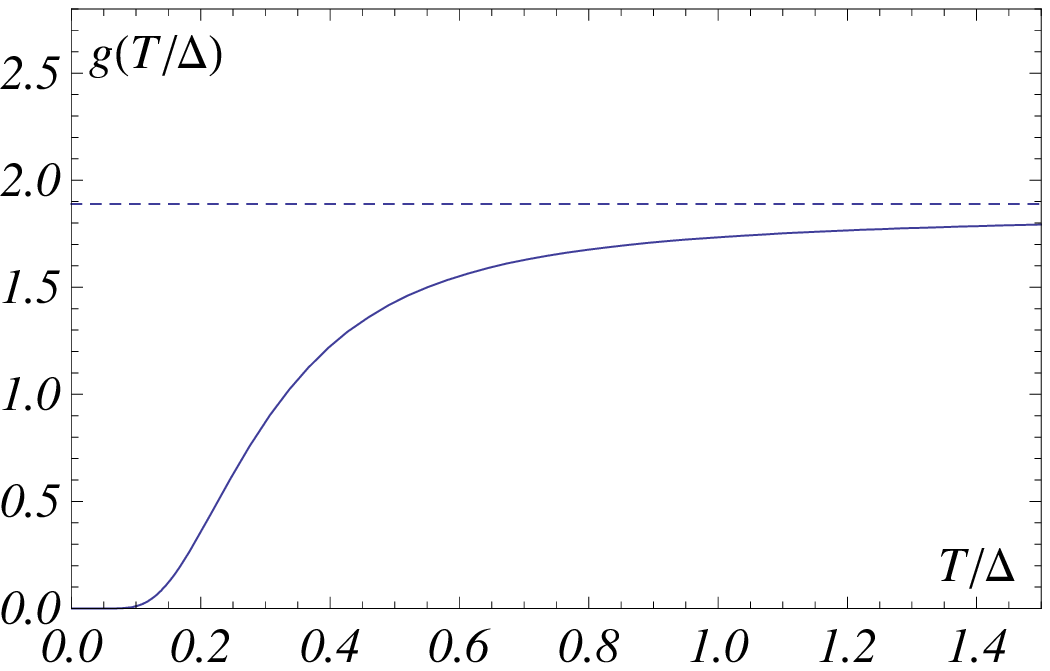}} \caption{\it
Top figure: function $f(t=T/|\Delta|)$ entering the expression of the specific heat
 (\ref{CVf}), in the metallic phase ($\Delta <0$). Bottom figure: function $g(t=T/\Delta)$
  entering the expression of the specific heat
 (\ref{CVg}), in the insulating phase ($\Delta >0$).} \label{fig.CV}
\end{figure}
The function $g(t)$ varies exponentially at small $t$, so that the specific interpolates from the $T^{3/2}$ behavior at the transition to an activated behavior.

\subsection{Landau quantization}

We now calculate the spectrum of the universal Hamiltonian in the presence of a magnetic field $B$.
 In the Landau gauge $\A=(0,Bx,0)$, the substitution $q_x \rightarrow q_x - e B y$ leads to the new Hamiltonian

$${\cal H}=  \left(
              \begin{array}{cc}
                0 & \Delta + {1 \over 2} m^* \omega_c^2 \tilde y^2 - i c_y q_y \\
               \Delta + {1 \over 2} m^* \omega_c^2 \tilde y^2 - i c_y q_y & 0\\
              \end{array}
            \right)
$$
where, as usual, $\tilde y= y - q_x/e B$. The spectrum in the bulk does not depend on the shift $q_x/eB$. We
 introduce the dimensionless variables $Y=    \tilde y/\alpha$ and $q_y =/\alpha
P$, so that $[X,P]=i$. We obtain the effective magnetic length for this problem $\alpha= \left({2 c_y
\over m^*
\omega_c^2 }\right)^{1/3}=(2 m^* c_y \ell_B^4)^{1/3}$, where $\ell=1/\sqrt{e B}$ is the usual magnetic length.
 The Hamiltonian is
rewritten as

\be {\cal H}= \left({m^* \omega_c^2 c_y^2 \over 2}\right)^{1/3} \left(
              \begin{array}{cc}
                0 & \delta + Y^2 - i P \\
                \delta + Y^2 + i P & 0\\
              \end{array}
            \right)
            \ . \label{HLandau}
\ee
Squaring this Hamiltonian, we have to solve the effective
Schr\"odinger equation
$$\ep_n^2 \psi^{A,B}=\left({m^* \omega_c^2 c_y^2 \over 2}\right)^{2/3} \big( P^2+(\delta
+ Y^2)^2 - i s [P,Y^2] \big)\ \psi^{A,B}$$
 where $s=\pm 1$
corresponds to the two sites $A$ and $B$. We have introduced the
dimensionless gap
\be \delta= {\Delta \over \left({ m^* \omega_c^2 c_y^2 \over
2}\right)^{1/3} } \propto {\Delta \over B^{2/3}}\label{deltadef} \ee

\noindent
We now have to diagonalize the effective hamiltonian ${\cal
H}_{eff}$  :

\be {\cal H}_{eff}=  P^2+(\delta
+ Y^2)^2 -  2 s Y \ , \label{Heff} \ee
and the eigenvalues $\ep_n$ of the original problem (\ref{HLandau})
are related to the eigenvalues $E_n$ of this effective Hamiltonian (\ref{Heff}) by

\be \ep_n = \pm \left({m^*\omega_c^2 c_y^2 \over 2}\right)^{1/3}\sqrt{E_n(\delta)} =\pm {\Delta \over \delta}\sqrt{E_n(\delta)} \, \label{deltaDelta}
 \ee
 where $\delta$
is given by (\ref{deltadef}). We thus obtain the general scaling behavior of the Landau levels.

\be
 \ep_n \propto B^{2/3} f_n(\Delta/B^{2/3})
\label{taugenW2} \ee

\begin{figure}[!h]
\centerline{ \epsfxsize 6cm \epsffile{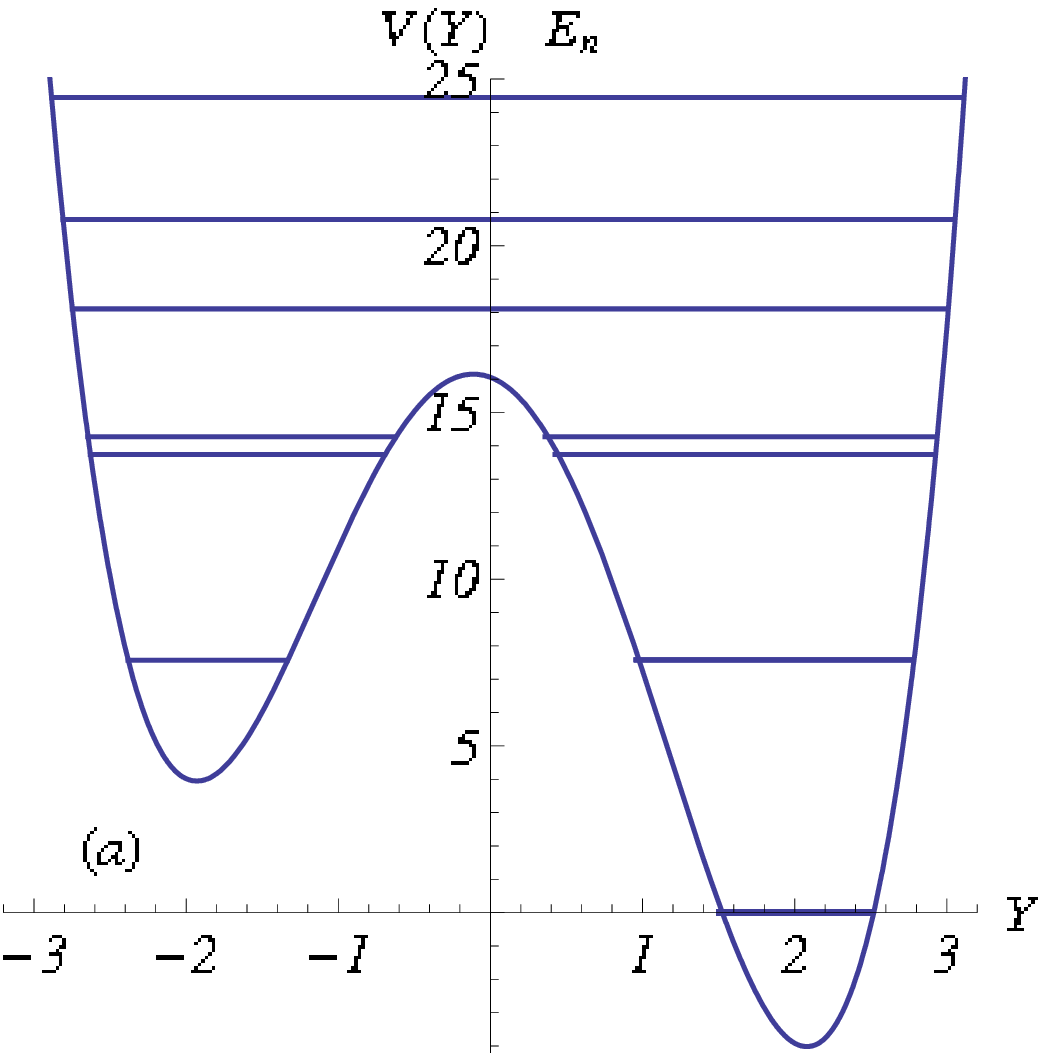} \epsfxsize 6cm
\epsffile{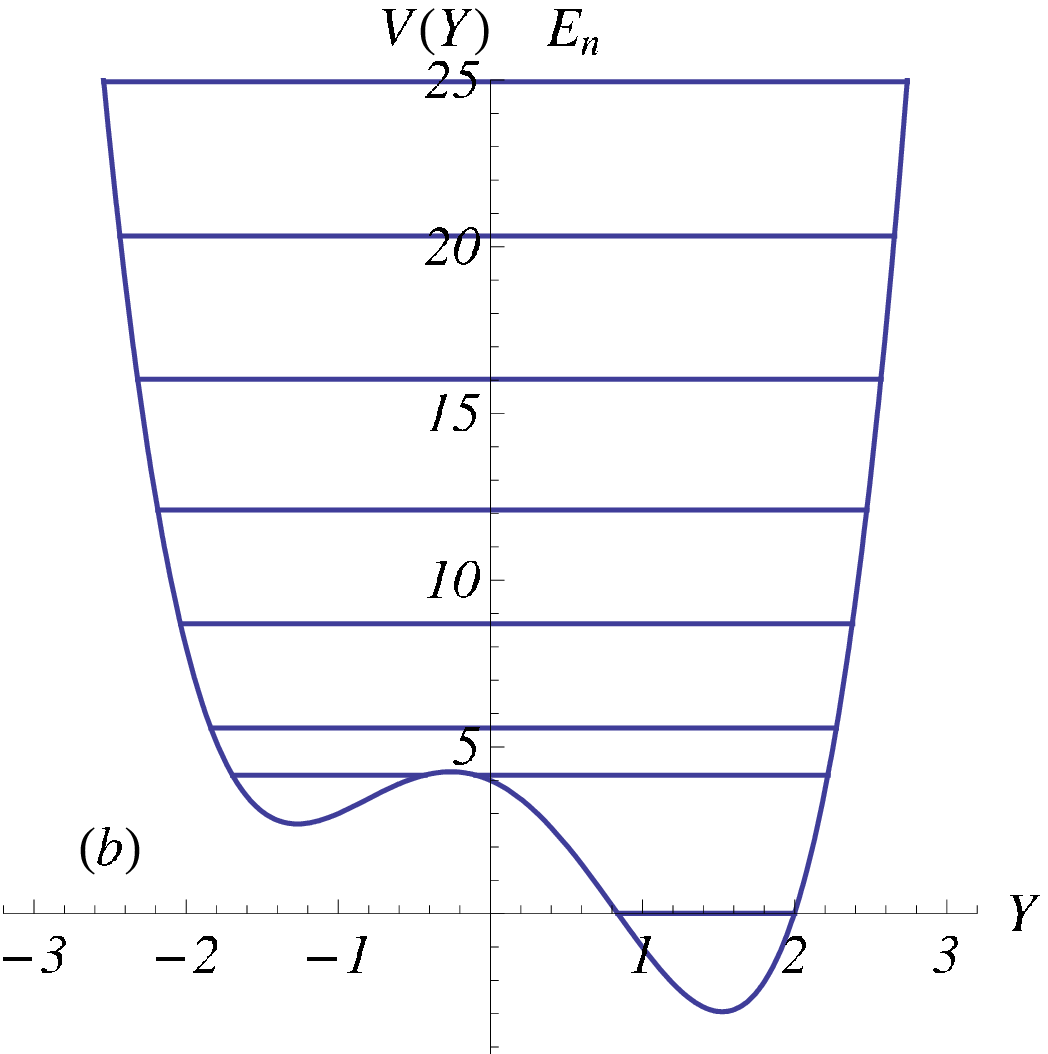}}
\centerline{ \epsfxsize 6cm \epsffile{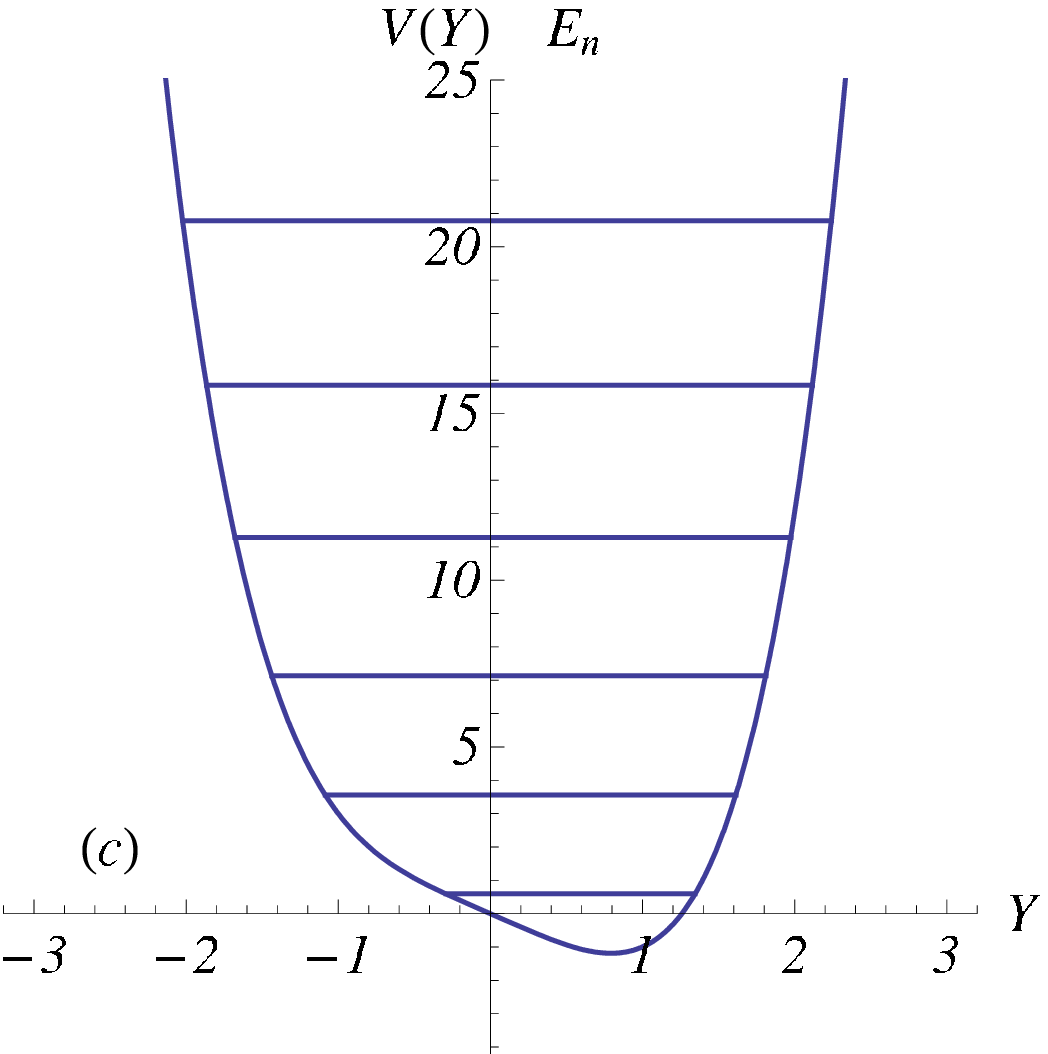} \epsfxsize 6cm
\epsffile{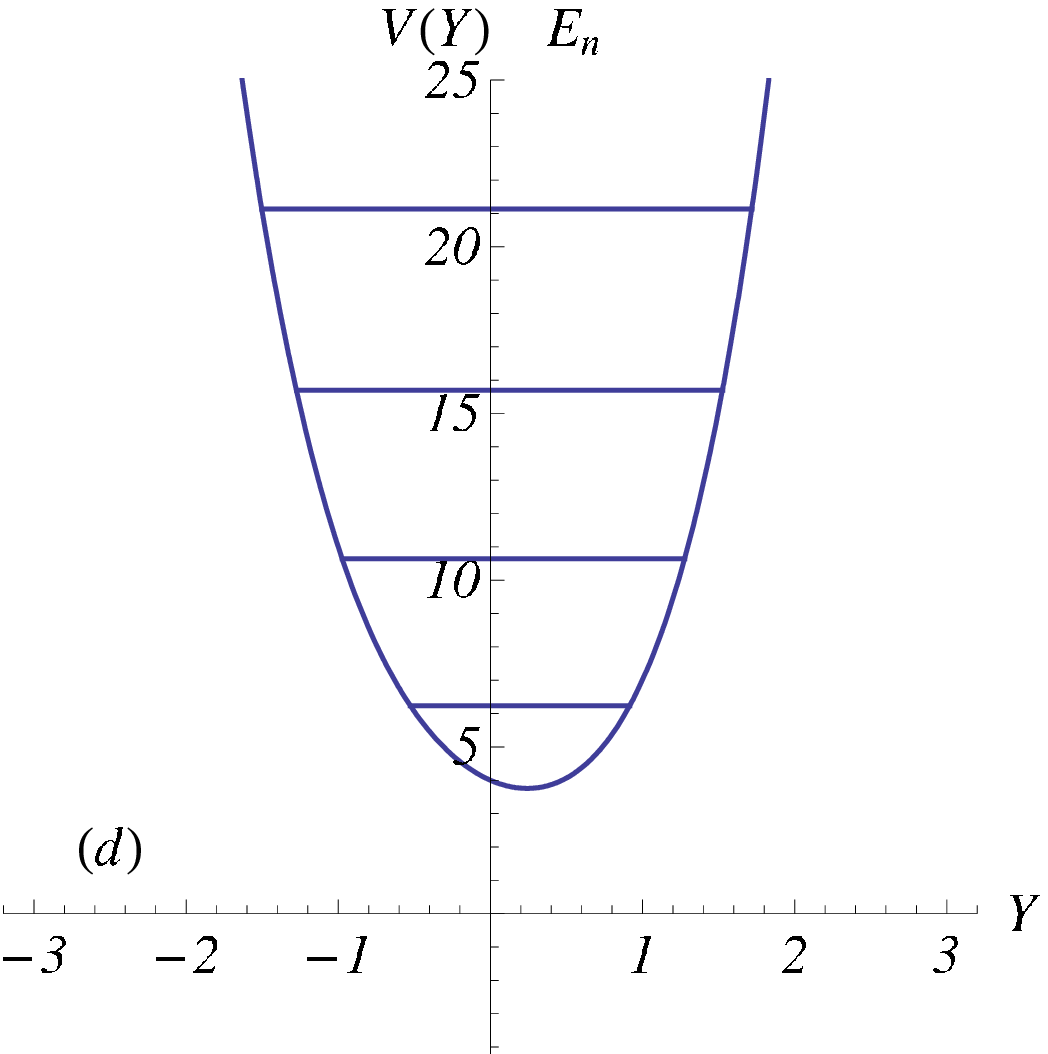}}
 \caption{\it Potential profile and energy
levels $E_n$  of the Hamiltonien ${\cal H}_{eff}= P^2
+ (\delta + Y^2)^2 - 2 Y$, for $\delta=-4,-2,0,2$. }
\label{fig.paraboles}
\end{figure}

When varying $\delta$, this Hamiltonian
has the remarkable property to describe continuously the Landau
level spectrum from the $\ep_n \propto \sqrt{n B}$ dependence with double degeneracy for
well separated  Dirac cones to the $\ep_n \propto (n+1/2)B$ usual
dependence for a massive particle.
The physics behind is that for  negative $\delta$, the problem is similar to the one of a particle in a
double well potential. In the limit of large negative $\delta$, that is far from the transition or in a weak
magnetic field, the potential has two well separated valleys which are almost uncoupled. This corresponds
to the situation of two independent valleys. Note that in this limit the energy shift between the two valleys is
$2 \sqrt{\delta}$. When $\delta$ diminishes, we progressively increase the coupling between valleys.
 The degeneracy of Landau levels is progressively lifted, as shown on Fig. \ref{fig.paraboles}.

 We have solved numerically the
Hamiltonian ${\cal H}_{eff}$. Eigenvalues are given on Fig.
(\ref{landautransition}) as functions of $\delta$. We now comment our results and the different limits.
\medskip

\begin{figure}[!h]
\centerline{ \epsfxsize 8cm \epsffile{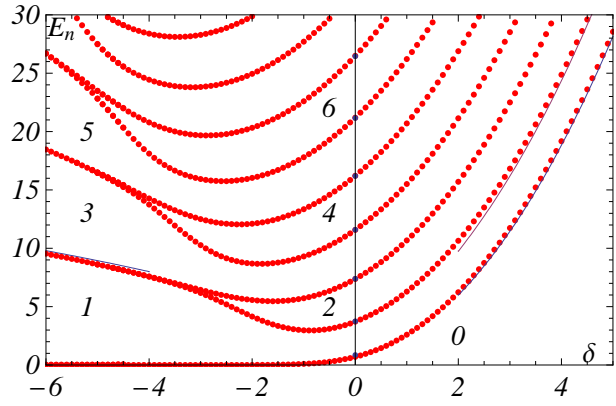}}
\caption{\it Energy levels $E_n(\delta)$  of the Hamiltonian ${\cal
H}_{eff}= P^2 + (\delta + Y^2)^2 - 2 Y$. We have plotted
asymptotic analytical behaviors, $E_n= 4 n \sqrt{-\delta}$ for large negative $\delta$
 and $E_n= \delta^2 + 4 (n+1/2) \sqrt{\delta}$ for large positive $\delta$.
.} \label{landautransition}
\end{figure}

\begin{figure}[!h]
\centerline{ \epsfxsize 8cm \epsffile{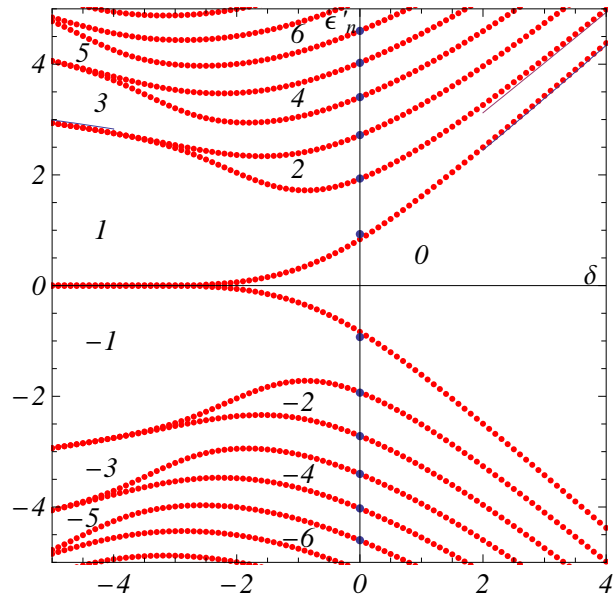}}
\caption{\it Energy levels $\ep_n'(\delta)= \ep_n(\delta)/(m^* \omega_c^2 c_y^2/2)^{1/3}$  as a function of the dimensionless gap $\delta$.
The dots on the $\delta=0$ axis indicate the semiclassical levels of the quartic Hamiltonian. }
\label{landautransition3}
\end{figure}

$\bullet$ If $\delta < 0$, we have to solve the problem of a double
well potential (Figure \ref{fig.paraboles}.a) (and independent wells in the limit $|\delta| \gg
1$). The potential has two minima for $Y_0= \pm \sqrt{|\delta|}$. In real space, the distance $2 Y_0$ corresponds
to the distance $2 y_0= 2 \alpha Y_0=2 q_D \ell_B^2$, where $\ell_B$ is the magnetic length.
An
expansion around these minima $Y=\pm \sqrt{\delta} +x$ gives the effective Hamiltonian :

$${\cal H}_{eff}=   P^2+4 |\delta| x^2 \mp 2  s \sqrt{|\delta|}  \ . $$
Introducing the new variables $p'=  P/ (\sqrt{2} |\delta|^{1/4})$ and
$x'=x \sqrt{2} |\delta|^{1/4}$, this Hamiltonian reduces to

$${\cal H}_{eff}=2 \sqrt{|\delta|}    ( p'^2+ x'^2  \pm 1 )$$
with eigenvalues

\be E_n= 4 n \sqrt{|\delta|} \ee Each level $n \neq 0$ is doubly
degenerate, due to the twofold structure of the potential well. We
deduce that
$$\ep_n^2 = 4 {\Delta^2 \over |\delta|^{3/2}} n $$
 which can be written in the usual form, introducing $c_x=\sqrt{- 2 \Delta /m^*}$

$$\ep_n = \pm \sqrt{2 n e c^2 B}  $$
with the velocity $c$ defined as
$$ c =\sqrt{ c_x c_y}$$
Each energy level $\ep_n$ is doubly degenerate. We recover the well-known result for two independent Dirac valleys, generalized here to the anisotropic case.

$\bullet$ When $|\delta|$ diminishes, the potential barrier between the two
valleys decreases and tunneling between the valleys  removes the
twofold degeneracy of each level (Figure \ref{fig.paraboles}.b).
 We can estimate the shift of the levels due to a finite
$\Delta$. The shift is proportional to the probability to tunnel between the two valleys. It scales as $\delta E_n \propto e^{-\sqrt{V-E_n} d}$
where the potential height $V$ is proportional to $|\delta|^2$ and the distance between valleys $d$ is proportional to $\sqrt{|\delta|}$.
As a result, the level degeneracy   is lifted as

\be e^{-|\delta|^{3/2}}  \sim e^{- \# {|\Delta|^{3/2} / B}}\ . \label{lift} \ee

\bigskip

$\bullet$ At the transition point, $\delta=0$, the energy levels are
those of a modified quartic oscillator with a potential $V(Y)=Y^4 - 2 Y$ and they have been obtained in Ref. \onlinecite{Dietl} and are well approximated by:
:

\be E_n=C (n+1/2)^{4/3} \ee
with $C= \pi^2 [3 \sqrt{2}/ \Gamma(1/4)^2]^{4/3}\simeq 2.185$. From eq. (\ref{deltaDelta}), we deduce the following dependence of the Landau levels

\be \ep_n= \pm A (m^* c_y^2)^{1/3} [ (n+1/2)\omega_c]^{4/3} \label{Dietl2}\ee
with $A= \sqrt{C/2^{3/2}}\simeq 1.173$. In ref. \onlinecite{Dietl}, we have studied in details the effect of the linear term in the
potential $V(Y)$, which only
slightly change the above result. We have attributed the phase term $1/2$ to the annihilation of the Berry phases attached to each Dirac point at their merging.
This is also briefly discussed in the next subsection of this paper.

\bigskip

$\bullet$ For large $\delta >0$, the Hamiltonian can be expanded  and transformed into a quadratic Hamiltonian
\be {\cal H}_{eff}= P^2 + \delta^2 + 2 \delta Y^2 - 2 Y \simeq
P^2 + \delta^2   + 2 \delta (Y - 1/(2 \delta))^2 \ee

so that the
spectrum is again the one of an harmonic oscillator
\be E_n=  \delta^2   + 2 \sqrt{2} (n'+1/2) \sqrt{\delta} \ee
and  we recover a usual Landau spectrum in the gapped phase.

\be \ep_n= \pm \big(\Delta +  \sqrt{{m^* c^2 \over  \Delta}} (n+1/2)
\omega_c \big) \ee

\subsection{Berry's phase}

We now briefly turn to the structure of the wave functions, solutions of the universal Hamiltonian (\ref{newH}).
They are of the form

\be \psi(\r)= {1 \over \sqrt{2}} \left(
                                   \begin{array}{c}
                                     1 \\
                                    e^{i \theta_{\q}} \\
                                   \end{array}
                                 \right)
                                e^{i \q \cdot \r}
                                 \ee
where the two components refer to the two sublattices $A$ and $B$.  The phase $\theta_{\q}$ is given by
 \be \tan \theta_{\q}= { c_y q_y \over \Delta + {q_x^2 \over 2 m^*}} \ .
 \ee
Note that the two valleys, centered on $q_x= \pm q_D= \pm \sqrt{- 2 m^* \Delta}$,  are described by the {\it same} wave function. The $\q$ dependence of $\theta_{\q}$  is shown on Fig. \ref{fig.berry} and exhibits a vortex structure around the two Dirac points.
Each  point is characterized by a Berry phase
${1 \over 2} \oint \nabla \theta_q \cdot d\q= \pm \pi$.
Fig. \ref{fig.berry} shows the  annihilation of the two Berry phases at the topological transition. This is the reason why the
 Landau levels acquire
a $n+1/2$ dependence near and above the transition  (see next section).\cite{Berryphase}

\begin{figure}[!h]
\centerline{ \epsfxsize 4cm \epsffile{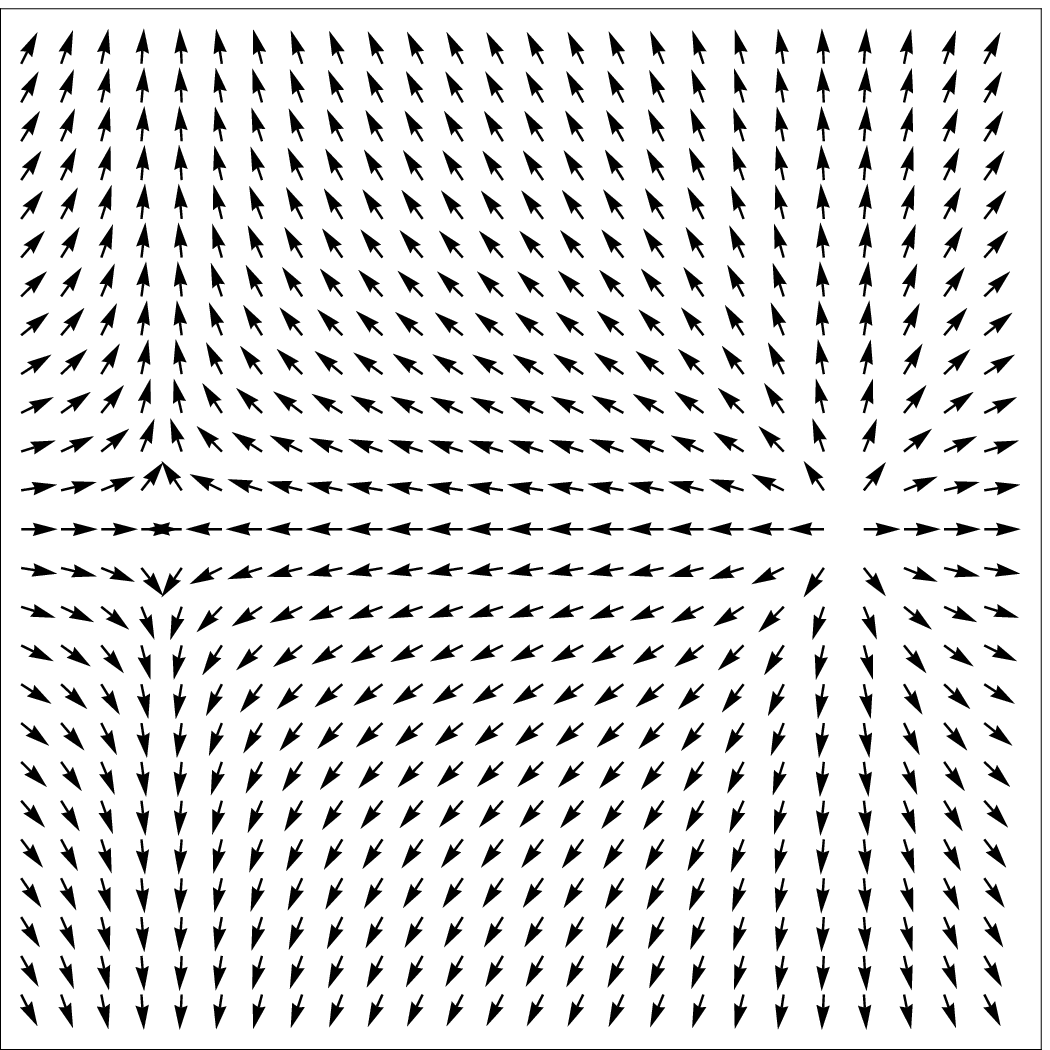} \epsfxsize 4cm
\epsffile{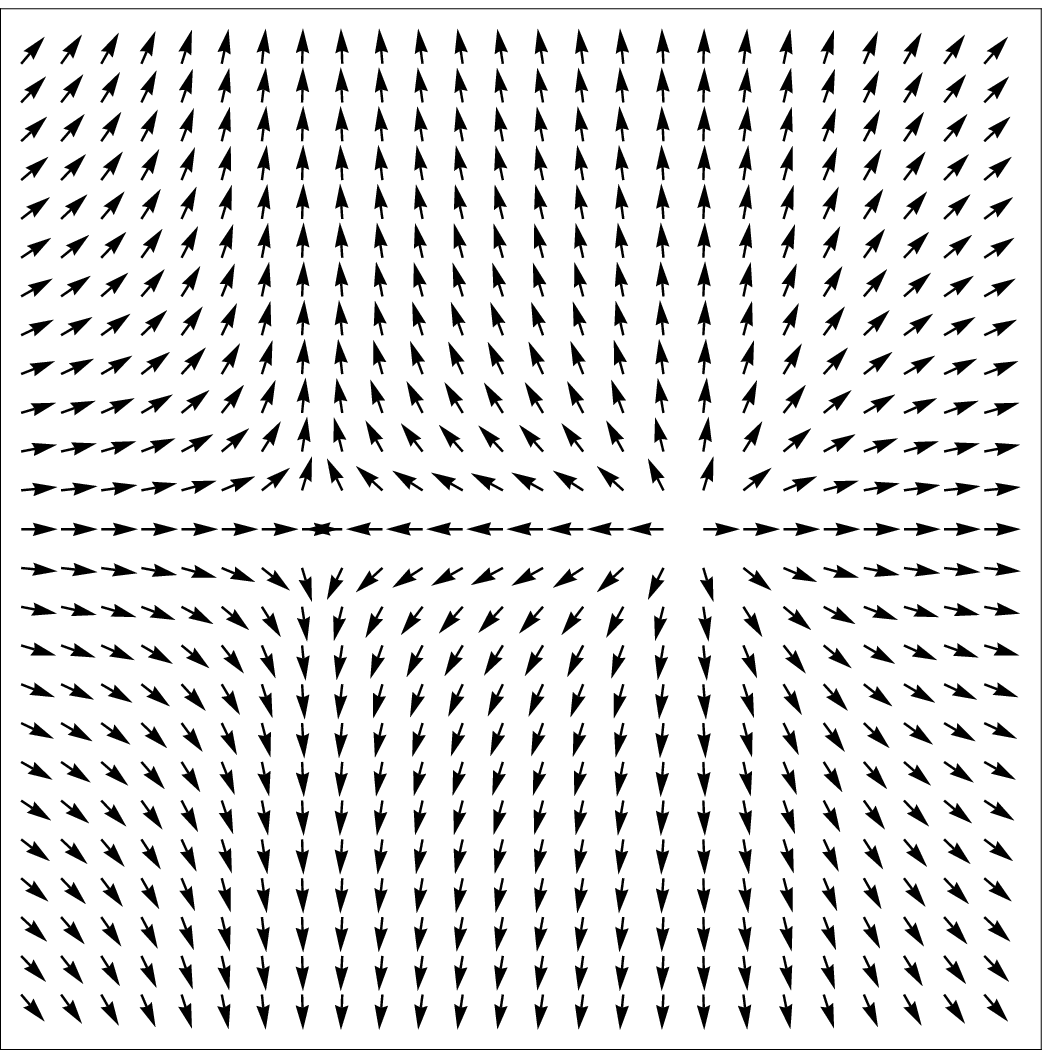}}
\centerline{ \epsfxsize 4cm \epsffile{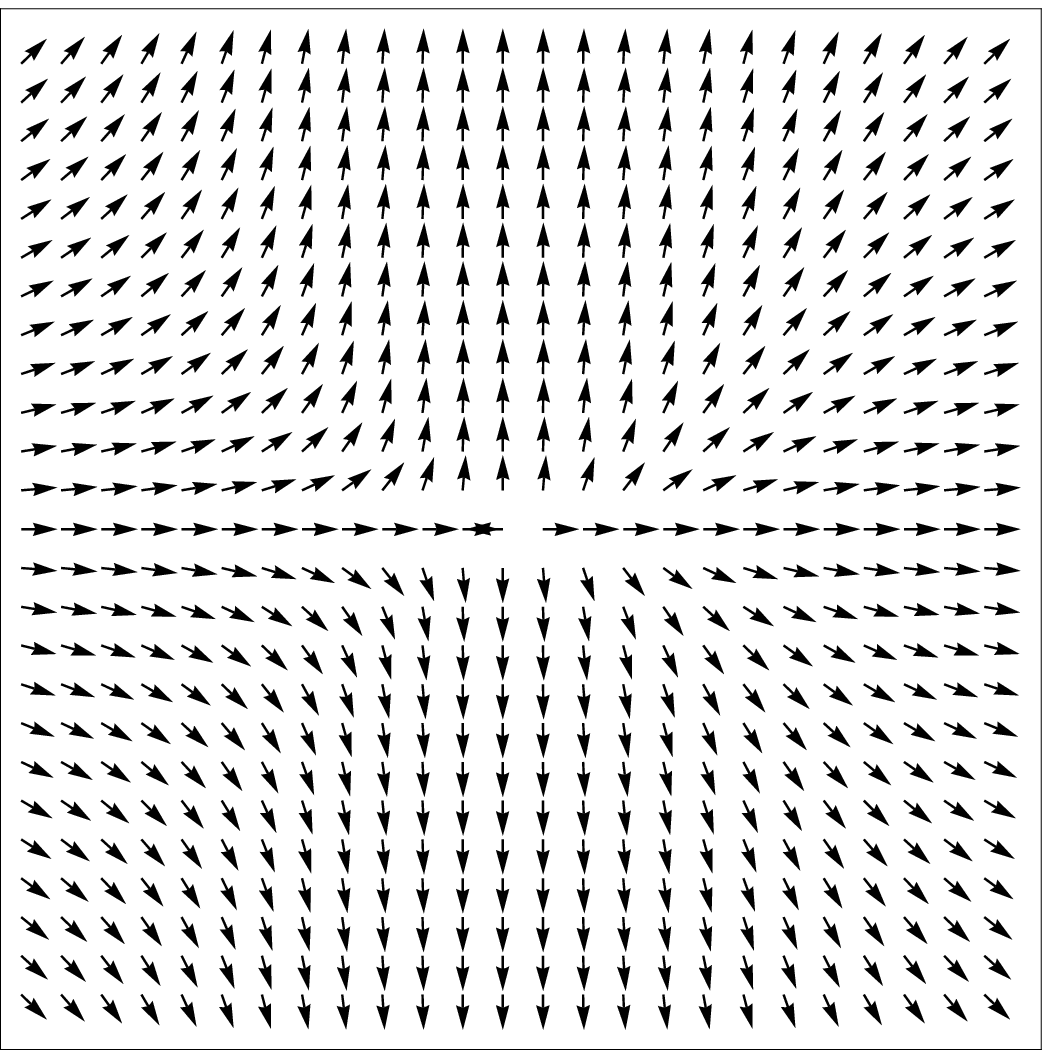} \epsfxsize 4cm
\epsffile{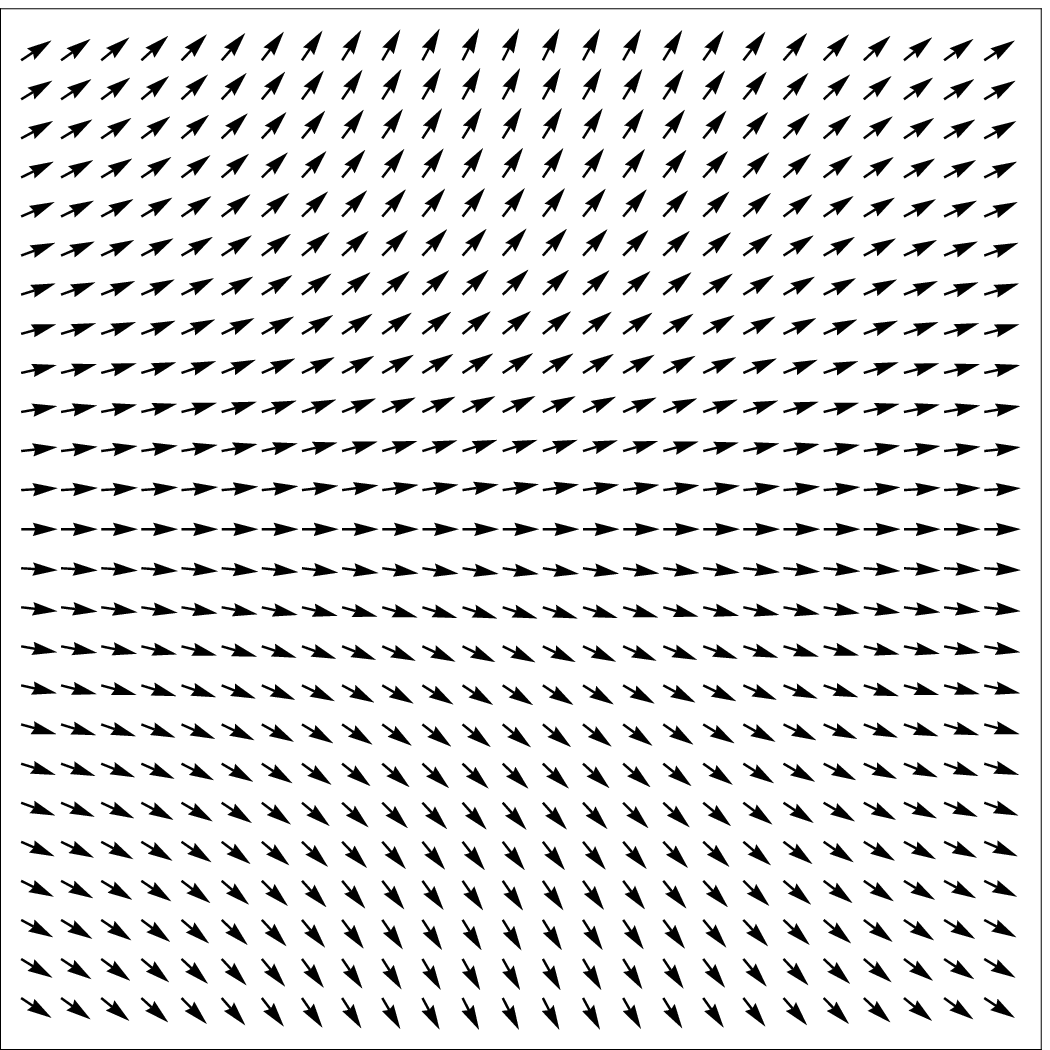}}
 \caption{\it Relative phase $\theta_{\q}$ of the  two-component wave function. The parameter are chosen in arbitrary units $m^*=c_y=1$. The four
 plots correspond respectively from left to right and then from top to bottom: $\Delta=-1, -.3, 0, 1$. In the insulating phase, two opposite Berry
 phases are attached to the two Dirac points. The Berry phases annihilate at the transition point. }
\label{fig.berry}
\end{figure}

\subsection{Semiclassical quantization and integrated density of states}

\begin{figure}[!h]
\centerline{ \epsfxsize 6cm \epsffile{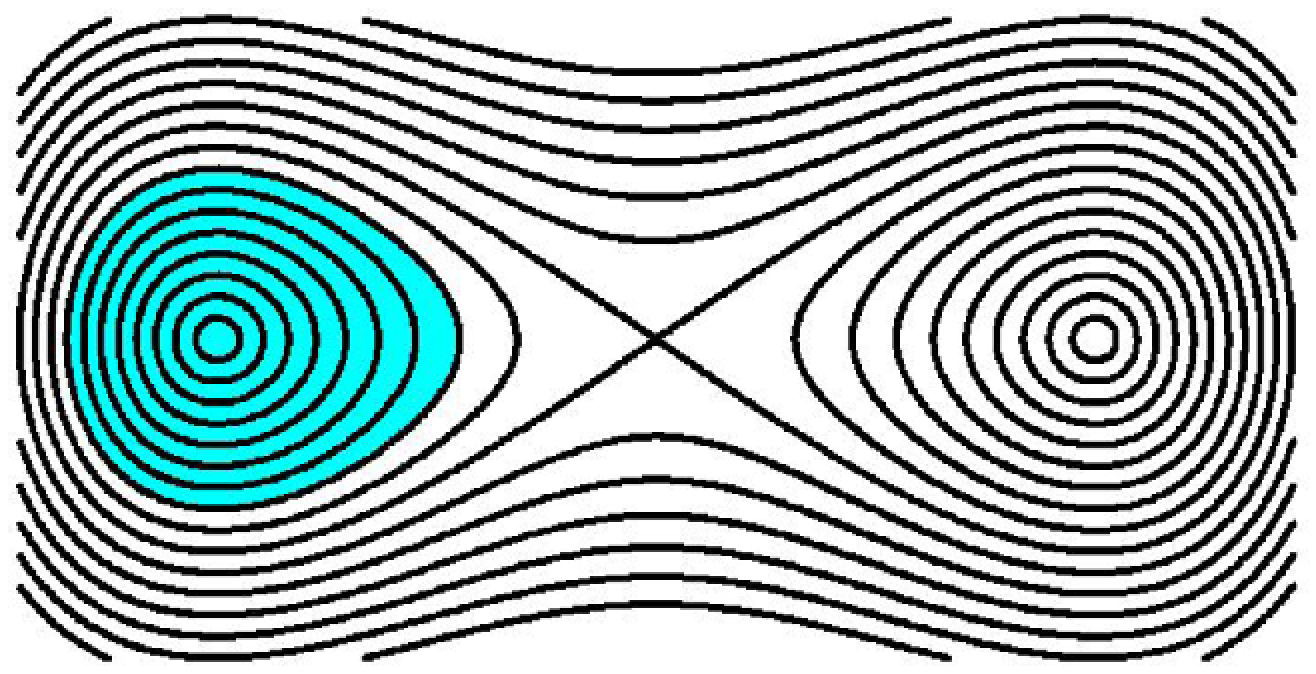}  \epsfxsize 6cm \epsffile{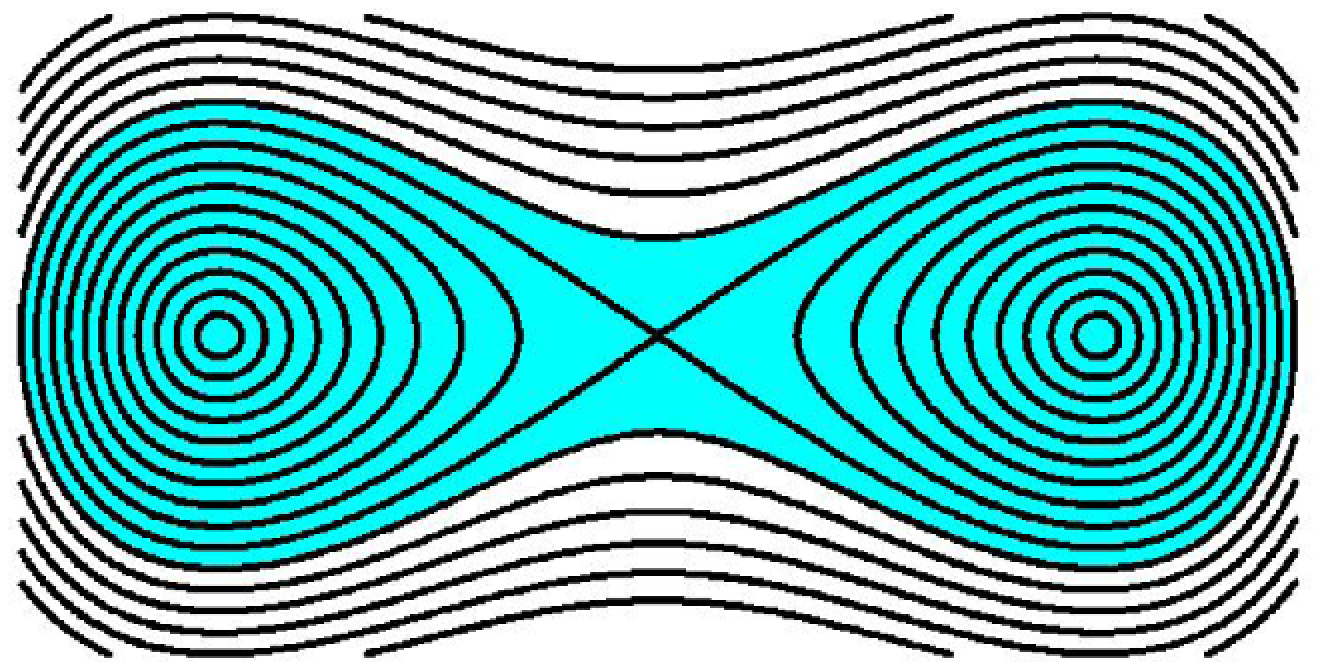} }
 \caption{\it Semiclassical quantization of area. When $\ep < -\Delta$, the quantization of energy levels results from the quantization  of orbits in each valley
$S(\ep) \ell_B^2 = 2 \pi n e B$ and the spectrum has the double valley degeneracy. When $\ep > -\Delta$, above the saddle point,
the quantization implies larger orbits which encircle the two Dirac valleys, and it reads $S(\ep) \ell_B^2 = 2 \pi (n'+1/2) e B$
 . }
\label{fig.areas}
\end{figure}

It is instructive to derive the energy levels from semiclassical Bohr-Sommerfeld quantization:
along one period of the motion, the action must be quantized. This condition
can be written as
\be {\cal{S}}(\ep)= 2 \pi (n + \gamma) {e B \over \hbar} \ , \ee
where ${\cal{S}}(\ep)$ is the area of a cyclotron orbit of energy $\ep$ is reciprocal space. It is simply ${\cal{S}}(\ep)= 4 \pi^2 N(\ep)$ where
$N(\ep)$ is the integrated density of states which can be obtained from expressions (\ref{DOS}).
The phase mismatch $\gamma$ is the sum of two contributions $\gamma=\gamma_M + \gamma_B$ where $\gamma_M$
is the Maslov contribution and $\gamma_B$ results from the Berry phase.
We obtain, for $\ep < -\Delta$:
\be {\cal{S}}(\ep)=  {4 \sqrt{2} \over 3} {\sqrt{ m^* (\ep - \Delta) } \over c_y}\left[(\ep + \Delta)
K\left(\sqrt{2 \ep \over \ep - \Delta }\right)- \Delta E\left(\sqrt{2 \ep \over \ep - \Delta}\right)\right] \label{Sepsbelow} \ee
where $K(x)$ and $E(x)$ are respectively  complete elliptic integrals of the first and of the second kind.\cite{gradstein}
This quantity represents the area enclosed by {\it each} of the two degenerate equal energy lines encircling one Dirac point (Fig. \ref{fig.areas}).  The phase mismatch cancels here due to
a finite Berry phase $\gamma_B=\pm 1/2$,\cite{Berryphase} so that the quantization condition is ${\cal{S}}(\ep)=2 \pi n e B$.

Similarly, for $\ep > -\Delta$:
\be {\cal{S}}(\ep)=   {8 \over 3}{ \sqrt{ m^* \ep } \over c_y}\left[(\ep + \Delta)
 K\left(\sqrt{\ep - \Delta \over 2 \ep}\right)- 2 \Delta E\left(\sqrt{\ep -
 \Delta \over 2 \ep}\right)\right] \label{Sepsabove} \ee
and the quantization condition is now ${\cal{S}}(\ep)=2 \pi (n'+1/2) e B$. The contribution $\gamma_B$ is
canceled since the semiclassical trajectories enclose the two Dirac points and the Berry phase is $0$.

Figure (\ref{fig.fit-semiclass}) compares the real spectrum with the above semiclassical quantization. The approximation works very well except in
the vicinity of the transition line $\ep_n =-\Delta$  which corresponds to the energy of the saddle point. It is worth stressing that the semiclassical
approximation describes perfectly well the vicinity of the topological transition (near $\delta=0$ axis on Fig. \ref{fig.fit-semiclass}.
The energy levels are given by the dimensionless equations

\begin{eqnarray}
\ep < - \Delta \qquad &\rightarrow& \qquad  F_-\left({\ep \over \Delta}\right)= {3 \pi \over 2} {n \over \delta^{3/2}} \label{QSCbelow} \\
\ep > - \Delta \qquad &\rightarrow& \qquad  F_+\left({\ep \over \Delta}\right)= {3 \pi \over 2 \sqrt{2} } {n'+1/2 \over \delta^{3/2}} \label{QSCabove}
\end{eqnarray}
with
\begin{eqnarray}
F_-(r)&=& \sqrt{r-1} \left[(r+1)K\left(\sqrt{{2 r \over  r-1}}\right) -  E\left(\sqrt{{2 r \over  r-1}}\right) \right] \label{QSC1} \\
F_+(r)&=& \sqrt{r} \left[(r+1)K\left(\sqrt{{r-1 \over 2 r}}\right) - 2 E\left(\sqrt{{r-1 \over 2 r}}\right)\right] \ .
\label{QSC2}
\end{eqnarray}

\begin{figure}[!h]
\centerline{ \epsfxsize 12cm \epsffile{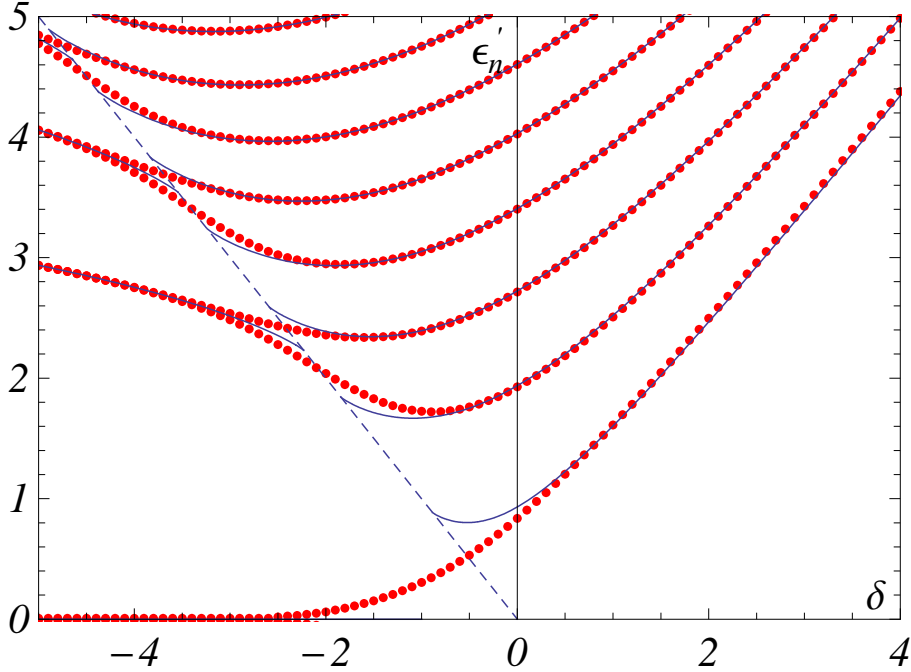}}
\caption{\it Positive exact energy levels $\ep_n'(\delta)= \ep_n(\delta)/(m^* \omega_c^2 c_y^2/2)^{1/3}$ (dots) compared
with the result of semiclassical quantization (lines). The dashed line $\ep_n'=-\delta$ corresponds to
$\ep_n=- \Delta$, that is to the energy of the saddle point. The discontinuity is due to the doubling
of the area $S(\ep)$ and to the cancelation of the Berry phase when energy crosses the saddle point.}
\label{fig.fit-semiclass}
\end{figure}

\section{Application to graphene and the honeycomb lattice}

We now propose that the effective Hamiltonian constitutes an
excellent description of the low energy physics of the $t-t'$ model including the two valleys. First we
briefly recall the electronic structure of graphene, and assume more
generally that one of the three hopping parameters $t'$ between
nearest neighbors may be different from the two others $t$, as shown
on Fig. (\ref{fig.reseau}).

\begin{figure}[!h]
\centerline{ \epsfxsize 6cm \epsffile{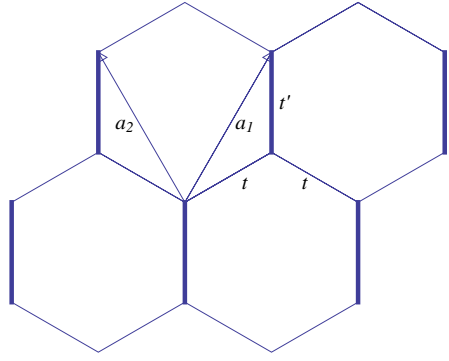} }
\caption{\it Honeycomb lattice with hopping integrals $t$ and $t'$,
and elementary vectors $\a_1$ and $\a_2$ discussed in the text.} \label{fig.reseau}
\end{figure}

The tight-binding Hamiltonian couples sites of different sublattices
named $A$ and $B$. The eigenvectors are Bloch waves of the form
\be |{\vec k} \rangle={1 \over \sqrt{N}} \sum_j \left(\, c_{{\vec
k}}^A |\R_j^A \rangle +c_{{\vec k}}^B |\R_j^B \rangle \, \right)
e^{i {{\vec k}}.\R_j} \label{bloch} \ee
where $|\R_j^A\rangle, |\R_j^B\rangle$ are atomic states. The sum
runs over vectors of the Bravais lattice. The Hamiltonian has the form (\ref{H}),
with
\be f({\vec k})= t' + t e^{i {\vec k} . \a_1} + t e^{i {\vec k} .
\a_2} \ee
where
$\a_1=a( {\sqrt{3} \over 2},{3  \over 2})$, $\a_2=a(- {\sqrt{3}
\over 2},{3 \over 2})$  are elementary  vectors of the Bravais
lattice, $a$ is the interatomic distance, and $t$, $t'$ are shown in
Fig. \ref{fig.reseau}. In Cartesian units
 \be f({\vec k})= t'+ 2 t \cos{\sqrt{3}\over 2}k_x a \  e^{i {3 \over 2} k_y a}
\label{fk} \ee
The energy, given by $\ep({\vec k})= \pm |f({\vec k})|$, is shown in
Figure (\ref{fig.isoenergy}) in the form of equal energy lines.

\begin{figure}[!h]
\centerline{ \epsfxsize 6cm \epsffile{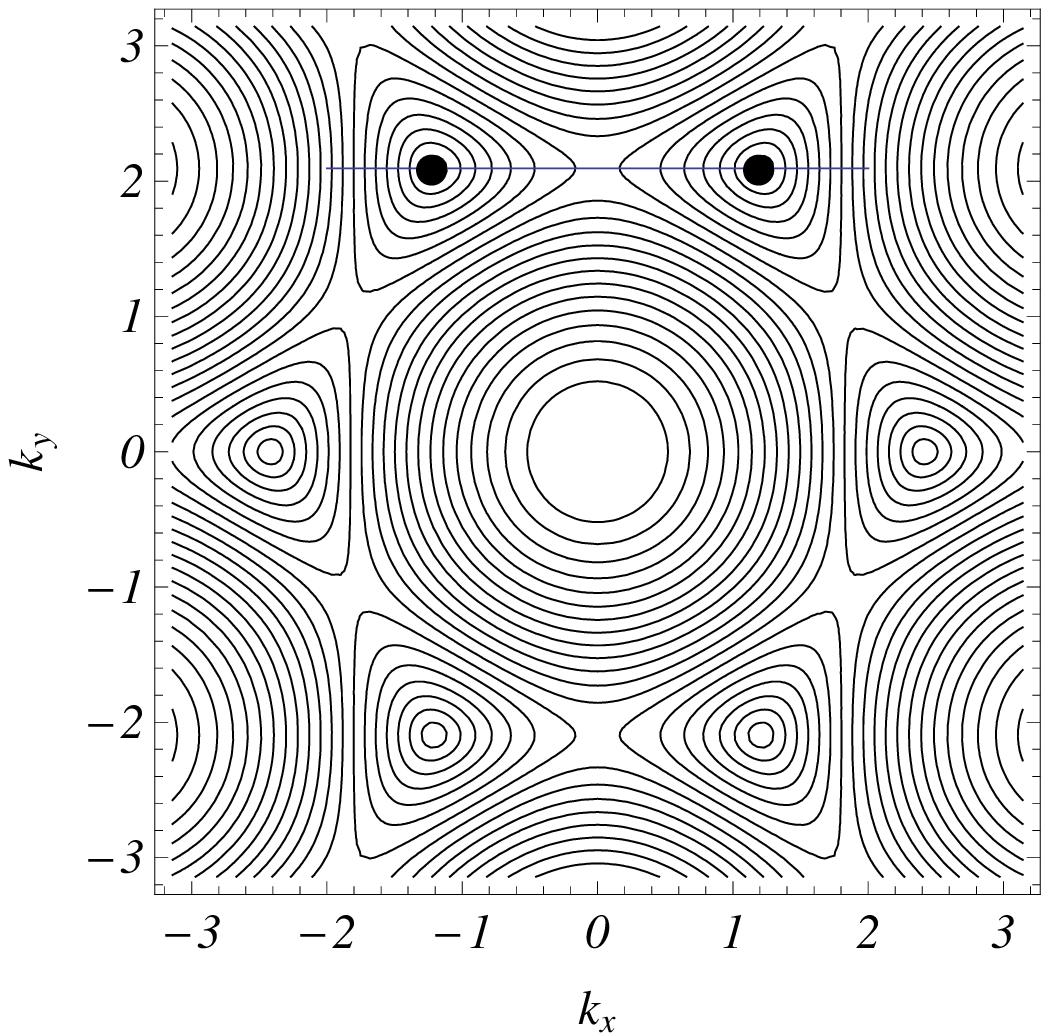} \epsfxsize 6cm \epsffile{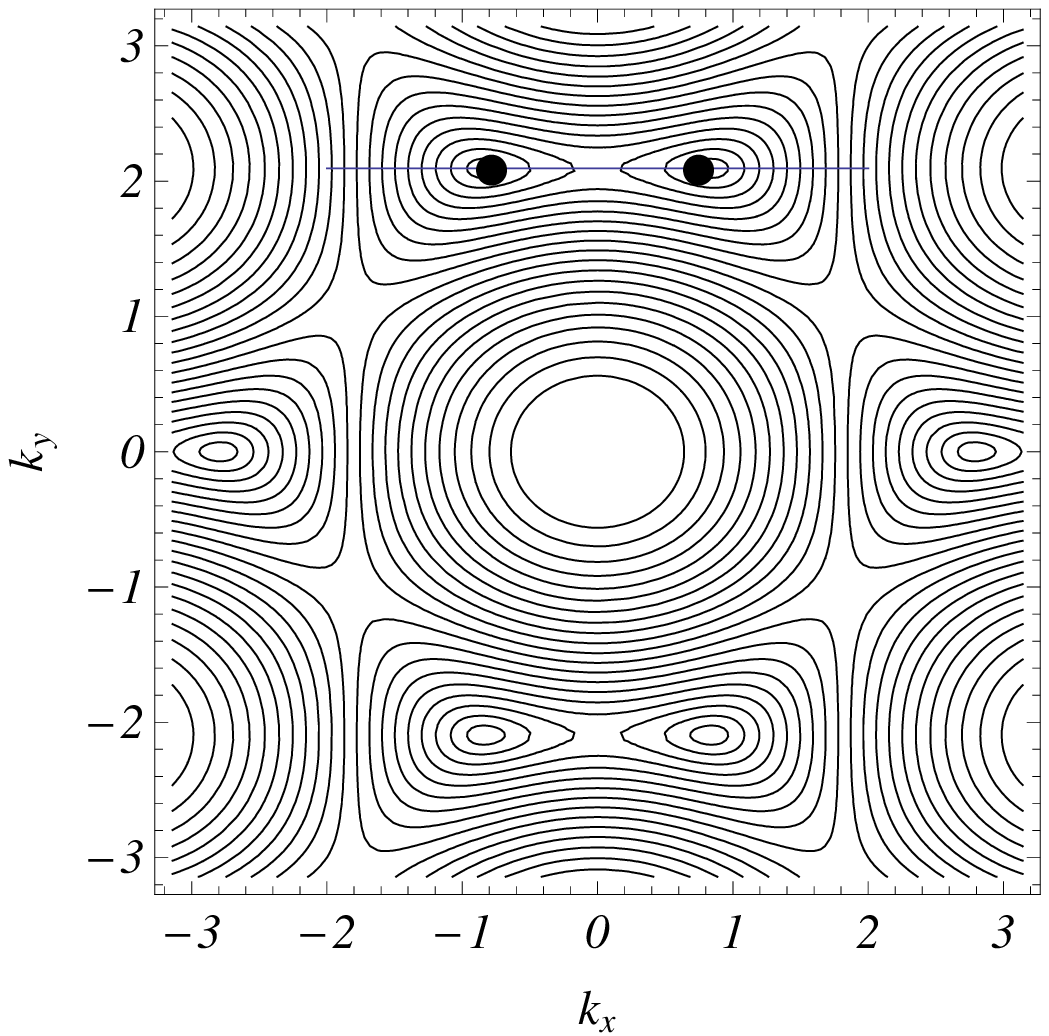}}
\centerline{ \epsfxsize 6cm  \epsffile{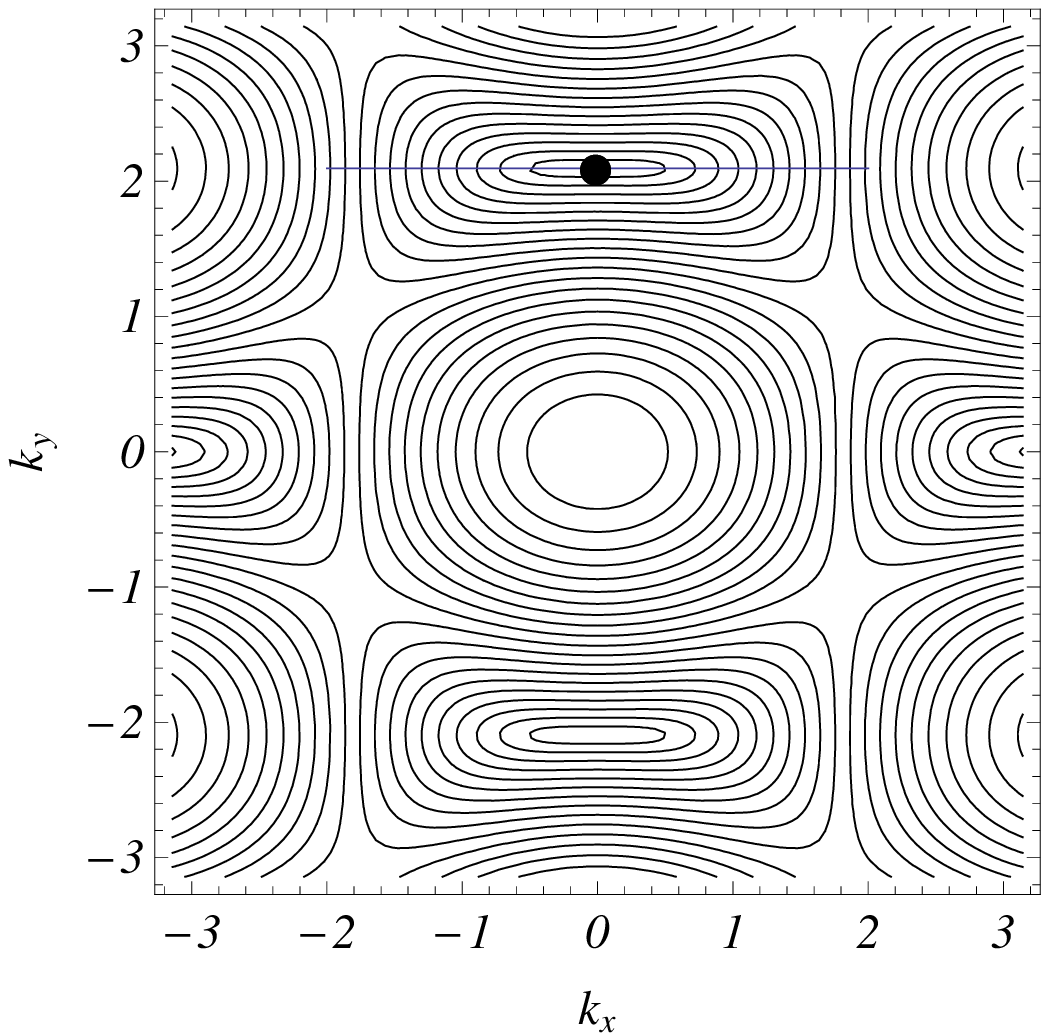}}
\caption{\it Isoenergy lines, for $t'/t=1,1.5,2$.} \label{fig.isoenergy}
\end{figure}

The evolution of the low energy spectrum when $t'$ varies is plotted on Fig. \ref{fig.bicone.graphene}. It is obviously well described
by our universal Hamiltonian (compare with Fig. \ref{fig.bicones}). In the following, we carefully map the two models on each other.
\begin{figure}[!h]
\centerline{ \epsfxsize 10cm \epsffile{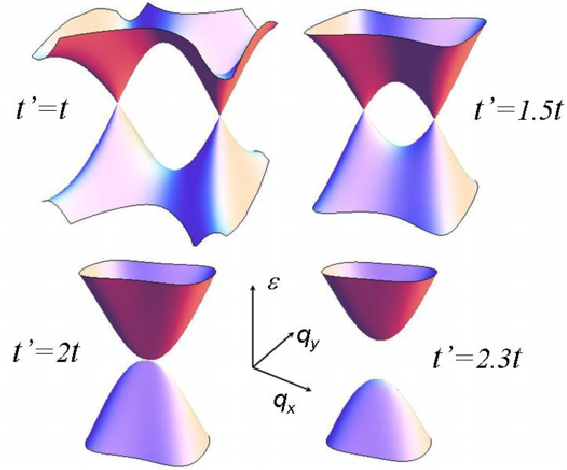}}
\caption{\it Evolution of the low energy spectrum when $t'$ approaches $t$. This evolution is very well described
by the universal Hamiltonian (see figure (\ref{fig.bicones}).} \label{fig.bicone.graphene}
\end{figure}

When $t'=t$, the energy vanishes at the two
points $\D$ and $\D'$  located at the corners $\K$ and $\K'$ of the
Brillouin zone $\K= 2 \a_1^*/3 + \a_2^*/3$, $\K'= \a_1^*/3 + 2
\a_2^*/3$, where $\a_1^*$ and $\a_2^*$ are reciprocal lattice
vectors) or, in Cartesians units
\be \D= \K= \left( {2\pi \over 3 \sqrt{3}
a},{2 \pi \over 3 a}\right) \qquad , \qquad \D'= \K'= \left( {-2\pi \over 3 \sqrt{3}
a},{2 \pi \over 3 a}\right) \ee

As $t'$ increases, the two points $\D$ and $\D'$ approach each
other. Their position is given by
 \be \D \ /  \D'= \left({ \pm { 2 \over 3 a} \arctan  \sqrt{{4 t^2 \over
 t'^2}-1},{2 \pi \over 3 a}}\right) \label{pos.D}\ee
They merge into the
single point $\D_0=(\a_1^*+\a_2^*)/2= (0,{2 \pi \over 3 a })$ when
$t'=2t$ (for $t'
> 2t$, a gap opens between the two subbands).
 ($a=1$  for shorter notations)

We now concentrate on the vicinity of the $\vec{KK'}$ axis ,
that is the line  $k_y={2 \pi/3}$. An expansion near this line, gives ($k_y=2 \pi/3 + q_y$):

\be  f({\vec k})= t' - 2 t \cos {\sqrt{3} \over 2} k_x  -3 i t q_y
\cos {\sqrt{3} \over 2}k_x \label{f1} \ee

We now wish to describe this Hamiltonian by the universal Hamiltonian (\ref{newH}), that is

 \be  f({\vec k})= \Delta - i c_y q_y + {q_x^2 \over 2 m^*}   \label{f2} \ee
for which we recall that fixing $\Delta$ and $m^*$ imposes the position $\pm q_D$ of the Dirac points and the velocity
 $c_x$ (see table \ref{table.parameters}).
We are now facing several possible choices   to properly introduce the effective Hamiltonian.
We may choose  to fix the mass $m^*$ and the parameter $\Delta$
by comparing the expansion of (\ref{f1})  near $q_x=0$:

\be  f({\vec k})= t' -  2 t + {3 \over 4} t q_x^2 - 3 i  t q_y
\ee
 with (\ref{f2}). This leads to
\be  \Delta= t' - 2 t \qquad , \qquad m^*= {2 \over 3 t} \qquad, \qquad c_y = 3 t               \ee
and $q_D$ and $c_x$ are obtained from table (\ref{table.parameters}) and are plotted in Fig. \ref{fig.fits}. This is not a good choice because, if it properly describes the spectrum near $q_x=0$, it does not
correctly describe the vicinity of the Dirac points $\pm q_D$.

We may also choose to fix $\Delta$ and the distance $2 q_D$ between the Dirac points and the velocity
$c_y$ around the Dirac points

\be \Delta = t'- 2 t  \qquad , \qquad  q_D= {2 \over \sqrt{3} } \arctan \sqrt{{4 t^2 \over t'^2} - 1} \qquad, \qquad
c_y = {3 \over 2 } t' \ee
so that the mass and the velocity $c_x$ are deduced from table (\ref{table.parameters}) and are plotted on Fig. \ref{fig.fits}. With this choice the velocity near the Dirac points
is not correct, so that the low energy spectrum when the Dirac points are far apart cannot be reproduced.

Among other possibilities we   finally  choose to fix $\Delta$ and the velocities $c_x$ and $c_y$. Comparing (\ref{f2}) with  the linear
expansion of (\ref{f1}) near the Dirac points
$$ f({\vec k})= {3 \over 2} i t' q_y   \pm
\sqrt{ 3(t^2 - {t'^2 / 4})} q_x $$
 where the  $\pm$ sign denotes the vicinity of the two points $\D$ and
$\D'$, we are led to  choose   the combination of parameters:

\be \Delta=t'-2 t \qquad, \qquad
 c_x = \sqrt{3} \sqrt{t^2 - t'^2/4} \qquad , \qquad
 c_y ={3 t' \over 2} \label{para1}
\ee
from which we deduce the effective mass

\be m^*= {- 2 \Delta \over c_x^2} = {8 \over 3 (2 t + t')} \label{para2} \ee
 This last choice of parameter is the best one since it properly describes the low energy spectrum with the correct velocities (see
  Fig. \ref{fig.fits}). Note that the low energy
 spectrum is not monotonic when $t'$ increases since the product $c^2=c_x c_y$ first increases and then decreases with $t'$ (figure \ref{fig.velocity}).

\begin{figure}[!h]
\centerline{ \epsfxsize 6cm \epsffile{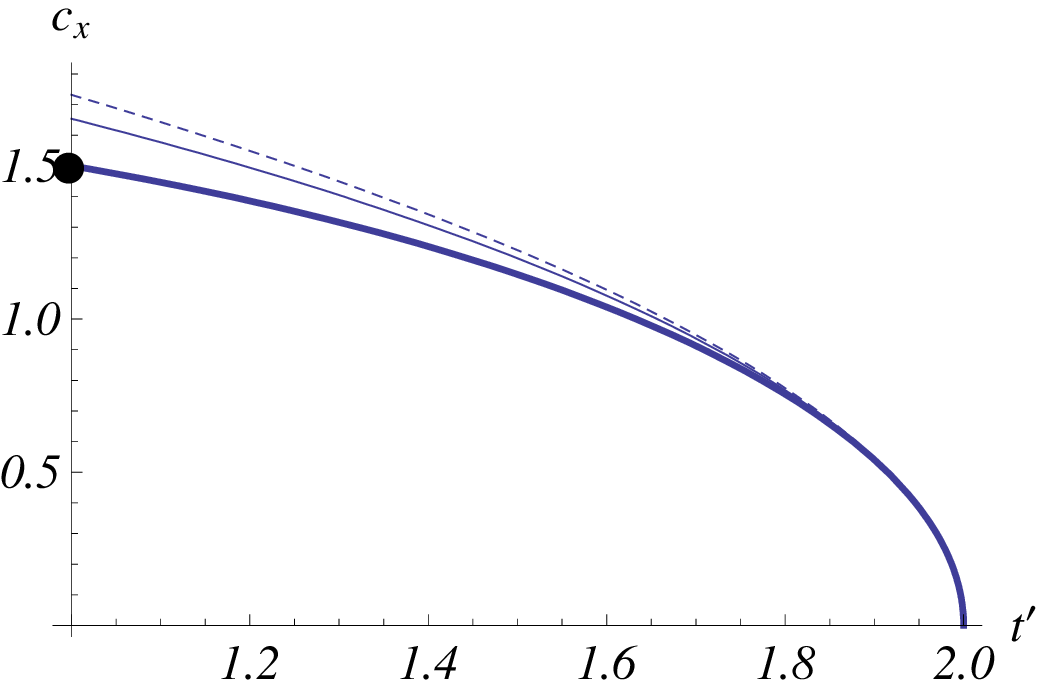} \epsfxsize 6cm \epsffile{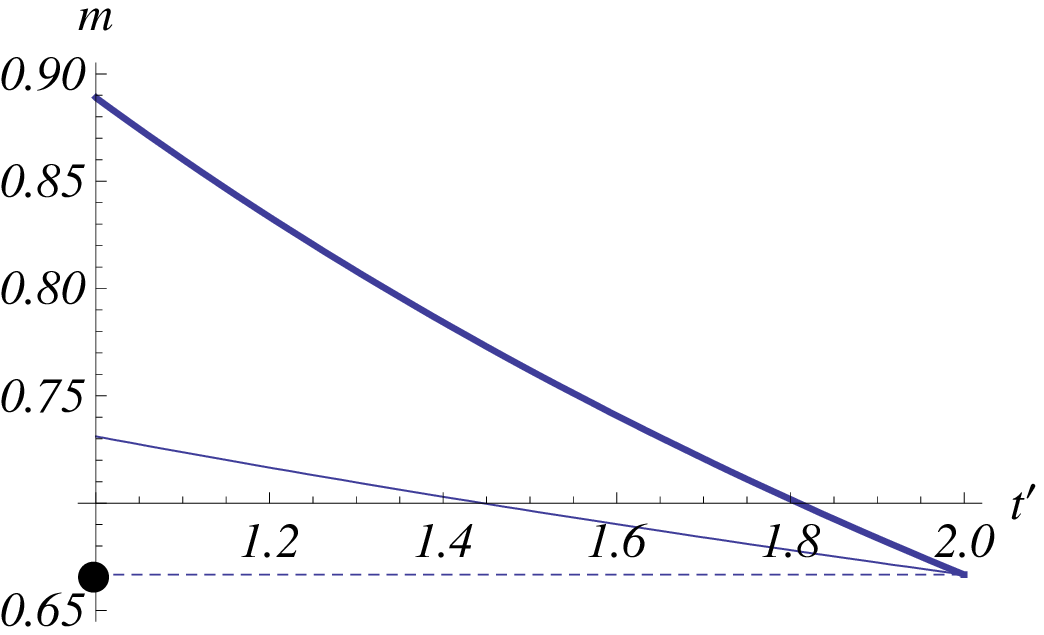}}
\centerline{ \epsfxsize 6cm  \epsffile{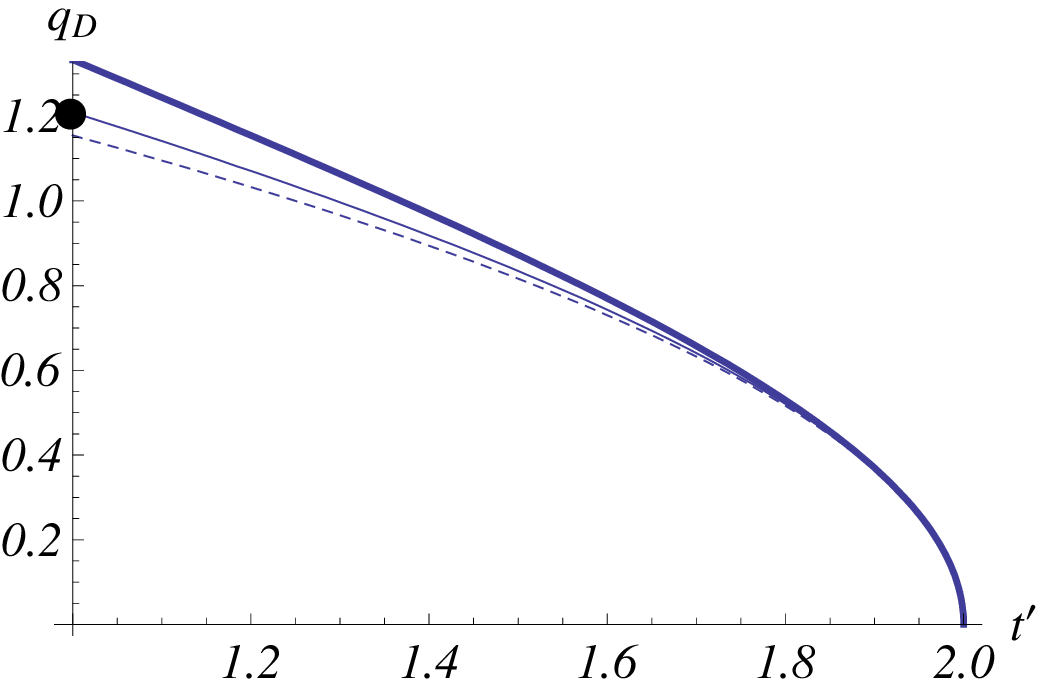}}
\caption{\it Plots  of the dependence of the quantities $c_x$, $m^*$ and $q_D$ as a function of $t'$.
The three lines correspond to three possible fits, where the mass $m^*$ is fixed (dashed lines), the position $q_D$ of the Dirac points is fixed (thin lines), or
the velocity $c_x$ is fixed, as chosen in the text (thick lines). The dot indicates the correct variation in the $t,t'$ model. } \label{fig.fits}
\end{figure}

\begin{figure}[!h]
\centerline{ \epsfxsize 8cm \epsffile{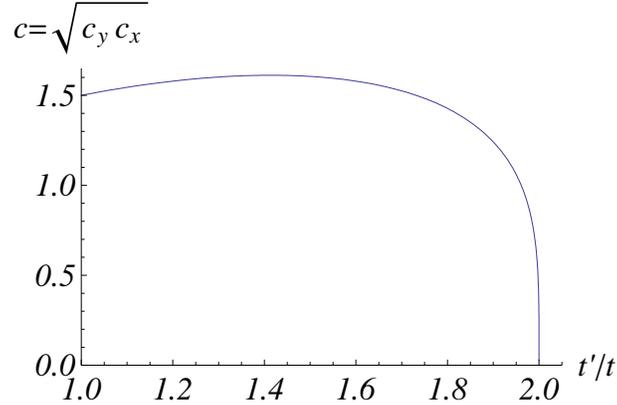} }
\caption{\it Non-monotonic behavior of the velocity $c=\sqrt{c_x c_y}= {3^{3/4} \over 2} \sqrt{t'} (4 t^2 - t'^2)^{1/4}$
 as a function of $t'/t$. The velocity is plotted in units of $t a/\hbar$.} \label{fig.velocity}
\end{figure}

Using these parameters, we can now apply the results of the universal Hamiltonian (figures \ref{landautransition}, \ref{landautransition3})
to the specific $t-t'$ model. We first introduce the   reduced flux
$f=\phi/\phi_0$, where $\phi$ is the flux through one elementary
cell of the honeycomb lattice, and $\phi_0=h/e=2 \pi /e$ is the flux quantum. We have $f= B
a^2 3 \sqrt{3}/(2 \phi_0)= {3 \sqrt{3} \over 4 \pi} e B$, since we
have chosen $a=1$, $\hbar=1$. From our study of the universal Hamiltonian, the energy levels are given by (\ref{deltaDelta})
\be \ep_n(f)= \pm {t'- 2 t \over \delta} \sqrt{E_n(\delta)} \ee
where the function $E_n(\delta)$ has been studied in section III.3 (figure \ref{landautransition}) and the parameter $\delta$ is
deduced from the parameters (\ref{para1}, \ref{para2}). We have

 \be \delta= \left( {2 \over \pi} \right)^{2/3} {t'-2 t \over [(2 t
 + t') t'^2]^{1/3}} {1 \over f^{2/3}} \ . \ee
In particular, in low field :

\be \ep_n = \pm \sqrt{2 n e c_x c_y B} = \sqrt{2 \pi n t' (4 t^2 -
t'^2)^{1/2}} \sqrt{f} \ . \label{enf} \ee

Fig. \ref{fig.fit.hasegawa1p5}  represents the energy levels for the honeycomb lattice with $t'=1.5 t$. The spectrum
 is represented as a function of the reduced flux
$f$. In low field, the levels e have a $\sqrt{n f}$ behavior. Then the degeneracy of the levels
 is lifted as predicted in eq. (\ref{lift}), that is
 $\Delta \ep \propto e^{- \# (2 t - t')^{3/2}/ f}$. The overall spectrum is quite well described by the semiclassical quantization
 rule explicited in section III.5.

\begin{figure}[!h]
\centerline{ \epsfxsize 12cm \epsffile{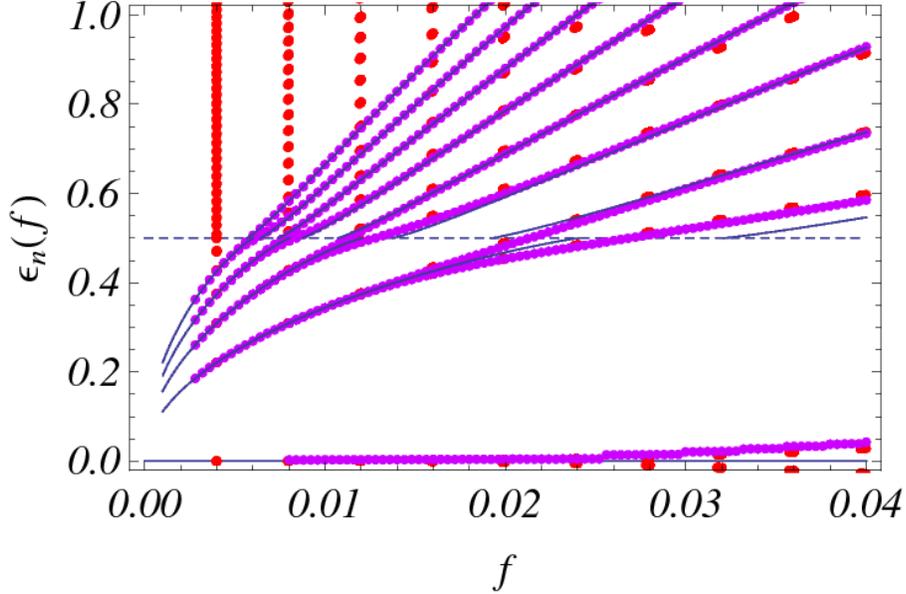}}
\caption{\it Color on line. Red dots : energy levels of the tight binding model on the honeycomb lattice
 with $t'=1.5 t$.\cite{Hasegawa2} Violet dots : energy levels calculated from the solutions of the 1D Schr\"odinger equation
  with a double
 well potential $V(Y)= (\delta - Y^2)^2 - 2 Y$. Full continuous lines : result of the semiclassical quantization rule. Dashed line :
  line $\ep= -\Delta= 2 t - t'$.}
   \label{fig.fit.hasegawa1p5}
\end{figure}

\section{summary}

We have shown that the motion and merging of Dirac points in a two-dimensional crystal can be fully described by a  simple $2 \times 2$
 Hamiltonian with a linear dispersion relation in one direction, a massive term in the other direction, and  gap term $\Delta$. By varying $\Delta$,
 a topological transition is driven, separating a semi-metallic phase with two Dirac points and a gapped phase.
 We have calculated analytically several quantities, such as the density of states, the specific heat, and the integrate density of states related to the area of semiclassical orbits. From this quantity, we obtain a simple semiclassical description of the Landau levels spectrum
 in a magnetic field $B$. More quantitatively, the problem in a magnetic field is related to
 a one-dimensional Schr\"odinger equation with a double well potential,
 whose potential barrier depends on the parameter $\Delta$ and the magnetic field $B$ as $\Delta/B^{3/2}$. The spectrum of Landau levels scales
 as $\ep_n \propto B^{2/3} f_n(\Delta/B^{3/2}) $. In the vicinity of the topological transition, it is very well described by the Bohr-Sommerfeld
 quantization rule $S(\ep) \ell_B^2 = 2 \pi (n+1/2)$.

 This Hamiltonian is appropriate to describe continuously the coupling between valleys which is usually neglected in the case of graphene, but which
 becomes important when approaching the topological transition. During completion of this paper,
 we have been aware of similar results in Ref.  \onlinecite{Kohmoto09}.


\begin{thebibliography}{99}
\bibitem{Wallace}   P.R. Wallace, Phys. Rev. {\bf 71}, 622 (1947)

\bibitem{review}  For a review see  A. H. Castro Neto, F. Guinea, N. M. R. Peres, K. S. Novoselov and  A. K. Geim,
  Rev. Mod. Phys. {\bf 81}, 109 (2009)

\bibitem{Hasegawa1} Y. Hasegawa, R. Konno, H. Nakano and M. Kohmoto,  Phys. Rev. B {\bf 74}, 033413 (2006)

\bibitem{Dietl}  P. Dietl, F. Pi\'echon and G. Montambaux, Phys. Rev. Lett. {\bf 100}, 236405 (2008)

\bibitem{CastroNeto2}  V. M. Pereira, A. H. Castro Neto and  N. M. R. Peres, arXiv.org/0811.4396

\bibitem{Guinea}  B. Wunsch, F. Guinea and  F. Sols, New J. Phys. {\bf 10}, 103027 (2008)

\bibitem{Montambaux09} G. Montambaux, F. Pi\'echon, J.-N. Fuchs and M.O. Goerbig, http://arxiv.org/abs/0904.2117

 \bibitem{Segev}O. Bahat-Treidel, O. Peleg, M. Grobman, N. Shapira, T. Pereg-Barnea, M. Segev, http://arxiv.org/abs/0905.4278

 \bibitem{Volovik} G.E. Volovik, Lect. Notes Phys. {\bf 718}, 31
(2007)


\bibitem{Katayama2006} S. Katayama , S. Kobayashi and Y. Suzumura, J. Phys. Soc. Jap. {\bf 75}, 054705 (2006)

\bibitem{Kobayashi2007} A. Kobayashi, S. Katayama, Y. Suzumura and H. Fukuyama,  J. Phys. Soc. Jap. {\bf 76}, 034711 (2007)

\bibitem{Goerbig2008} M.O. Goerbig, J.N. Fuchs, F. Pi\'echon and G.
Montambaux, Phys. Rev. B {\bf 78}, 045415 (2008)

\bibitem{Zhu} S.-L. Zhu, B. Wang and L.-M. Duan, Phys. Rev. Lett.
{\bf 98}, 260402 (2007)


\bibitem{Zhao}  E. Zhao and A. Paramekanti,  Phys. Rev. Lett. {\bf  97}, 230404 (2006).



\bibitem{Hou2009}  J.-M. Hou, W.-X. Yang and  X.-J. Liu, Phys. Rev. A {\bf 79}, 043621 (2009)

\bibitem{Lee09} K.L. Lee, B. Gremaud, R. Han, B.-G. Englert and C. Miniatura, http://arxiv.org/abs/0906.4158

\bibitem{McClure} J.W. McClure, Phys. Rev. {\bf 104}, 666 (1956)

\bibitem{Hofstadter} D. Hofstadter, Phys. Rev. B {\bf 14}, 2239 (1976)

\bibitem{Rammal} R. Rammal, J.  Physique {\bf 46}, 1345 (1985)

\bibitem{Hasegawa2} Y. Hasegawa and M. Kohmoto, Phys. Rev. B {\bf 74}, 155415 (2006)

\bibitem{Banerjee}  S. Banerjee, R. R. P. Singh, V. Pardo and  W. E. Pickett, Phys. Rev. Lett. {\bf 103}, 016402 (2009)

\bibitem{remark2} We consider here the case where there is only one pair of Dirac points, although
the number of pairs may be larger than one: F. Pi\'echon {\it et al.} in preparation

\bibitem{directions} We stress that the $x-y$ directions define local axes which are fixed by the band parameters $t_{mn}$.

\bibitem{gradstein} I.S. Gradshteyn, I.M. Ryzhik and A. Jeffrey, Tables of integrals, series, and products (Academic Press 2007)

\bibitem{Berryphase} G.P. Mikitik and Yu. V. Sharlai, Phys. Rev. Lett. {bf 82}, 2147 (1999)


\bibitem{Kohmoto09}  K. Esaki, M. Sato, M. Kohmoto and  B. I. Halperin, http://arxiv.org/abs/0906.5027




\end{thebibliography}
\end{document}